\documentclass[default]{aastex701}
\usepackage{siunitx}
\DeclareSIUnit{\pixel}{pixel}
\usepackage[version=4]{mhchem}
\usepackage{CJK}

\begin{document}
\begin{CJK*}{UTF8}{gbsn}

\title{Spectral Evolution of Ceres' Surface and Implications for Space Weathering}

\correspondingauthor{Jian-Yang Li}

\author[orcid=0000-0003-2263-5165,sname=Zhang,gname=Qing]{Qing Zhang (张庆)}
\affiliation{Planetary Environmental and Astrobiological Research Laboratory (PEARL), School of Atmospheric Sciences, Sun Yat-sen University, Zhuhai, China}
\email{zhangq735@mail.sysu.edu.cn}

\author[orcid=0000-0003-3841-9977,sname=Li,gname=Jian-Yang]{Jian-Yang Li (李荐扬)}
\affiliation{Planetary Environmental and Astrobiological Research Laboratory (PEARL), School of Atmospheric Sciences, Sun Yat-sen University, Zhuhai, China}
\affiliation{Xinjiang Astronomical Observatory, Chinese Academy of Sciences, Urumqi, China}
\email[show]{lijianyang@mail.sysu.edu.cn}

\author[orcid=0000-0002-9848-4650,sname=Yang,gname=Yazhou]{Yazhou Yang (杨亚洲)}
\affiliation{State Key Laboratory of Solar Activity and Space Weather, National Space Science Center, Chinese Academy of Sciences, Beijing, China}
\email{yangyazhou@nssc.ac.cn}

\author[orcid=0009-0001-5533-5671,sname=Liu,gname=Minge]{Minge Liu (刘民歌)}
\affiliation{State Key Laboratory of Solar Activity and Space Weather, National Space Science Center, Chinese Academy of Sciences, Beijing, China}
\affiliation{College of Earth and Planetary Sciences, University of Chinese Academy of Science, Beijing, China}
\email{liuminge23@mails.ucas.ac.cn}

\author[orcid=0000-0002-7699-2507,sname=Liu,gname=Yang]{Yang Liu (刘洋)}
\affiliation{State Key Laboratory of Solar Activity and Space Weather, National Space Science Center, Chinese Academy of Sciences, Beijing, China}
\affiliation{College of Earth and Planetary Sciences, University of Chinese Academy of Science, Beijing, China}
\email{yangliu@nssc.ac.cn}

\begin{abstract}

The surface of Ceres exposed to the space environment will be gradually altered, but the evolution of its spectral properties remains poorly understood. Here we analyze visible and infrared spectra acquired by the Dawn mission to investigate spectral variations across different surface units. \added{We identify local shifts of a few \si{\nm} toward shorter wavelengths in the \SI{2.7}{\um} band center, predominantly} associated with freshly exposed slumping and impact materials. Our analysis suggests that compositional variations in Ceres’ regolith are unlikely the explanation for the band center shift, and therefore it may represent a spectral evolution within \num{e6} years. Combining previous studies about the spectral reddening on Ceres, we propose a two-stage spectral evolution on Ceres. \added{The blue-shifted regions are interpreted as the youngest materials. Within \num{e6} years after exposure, these materials become spectrally bluer, the \SI{2.7}{\um} band weakens, and its center shifts toward longer wavelengths.} Then the evolution is characterized by spectral reddening and strengthening of the \SI{2.7}{\um} band without the band center shift. Based on the previous laboratory work and observations on other C-complex asteroids, we suggest that the observed spectral variations may be attributed to the chemical and/or physical changes of the regolith, possibly associated with space weathering.

\end{abstract}

\keywords{\uat{Asteroid surfaces}{2209} --- \uat{Spectroscopy}{1558}}


\section{Introduction} 

Dwarf planet (1) Ceres, the largest body in the asteroid belt, is classified as a C-class asteroid \citep{2009Icar..202..160D, 2022A&A...665A..26M}. Its surface is covered by ubiquitous Mg-phyllosilicates and ammoniated clays, mixed with Mg-Ca carbonates and dark components \citep{2015Natur.528..241D, 2018M&PS...53.1844D, 2016Sci...353.4279A, 2018SciA....4.1645C}, and probably a considerable fraction of carbonaceous chondritic (CC) materials accumulated from asteroid infall \citep{2019NatAs...3..140M, 2020JGRE..12506606K}. Water ice, organics, and salts are also identified in some local areas \citep{2016Sci...353.3010C, 2017Sci...355..719D, 2020NatAs...4..786D}. The surface of Ceres, exposed to the space environment, is expected to be gradually altered by space weathering. This process results in many physical and chemical changes to the surface, modifying the spectral characteristics \citep{2000M&PS...35.1101P,2001JGR...10610039H}. Ceres' surface composition is unique among inner solar system objects for which we have studied the spectral effects of space weathering \citep{2016JGRE..121.1865P}. The study of the spectral evolution on Ceres not only helps us understand its regolith composition and evolution, but also expands our understanding of spectral evolution on the C-complex asteroids in general. 

Laboratory simulations of space weathering \citep[e.g.,][]{2014Icar..237..278B, 2015AA...577A..41L, 2017Icar..285...43L} and analysis of remote sensing observations \citep[e.g.,][]{2020Sci...370.3660D, 2020Sci...368..654M} are two primary approaches to understanding the spectral evolution due to space weathering. Laboratory experiments on carbonaceous materials \citep[see reviews by, e.g.,][]{2023Icar..40015563C} have revealed that the spectral effects of space weathering on C-complex asteroids are diverse and complicated, depending on the surface albedo/composition and the weathering agents \citep[e.g.,][]{2017Icar..285...43L}. However, no meteorites in the existing meteorite collections have been identified as originating from Ceres \citep{2018M&PS...53.1793M}. The composition of Ceres is broadly consistent with CM/CI chondrite bulk composition, but CM/CI chondrites lack the \ce{NH4+} spectral feature at \SI{3.1}{\um}. The complex composition of Ceres could therefore lead to different spectral effects compared to those of the CM/CI chondrites.

The Dawn spacecraft conducted a global survey of Ceres, providing high spatial and spectral resolution data on a global scale and allowing for a comprehensive study of its spectral evolution. Results from the Dawn mission revealed that the geologically young units on Ceres generally feature higher albedos, spectrally blue colors, and slightly weaker hydroxyl band at \SI{2.7}{\um} and \ce{NH4+} band at \SI{3.1}{\um} than old terrains, suggesting possible reddening, darkening, and band weakening effects of space weathering \citep{2016P&SS..134..122N, 2016GeoRL..4311987S, 2019Icar..318...56S}. However, many questions remain. For example, the \num{2.7} and \SI{3.1}{\um} band depth variations have been attributed to the variations in the abundance of phyllosilicates \citep{2016Sci...353.4279A}, which cannot explain their correlation with surface age. Also, the mechanism causing the spectral slope variations on Ceres remains a subject of discussion \citep{2017GeoRL..44.1660S, 2021NatCo..12..274S}. Furthermore, in terms of timescales, the youngest units on Ceres identified by crater counting are about a few million years old \citep{2019Icar..318...56S}, which is comparable to the timescale of space weathering on S-type asteroids in the main belt \citep{2009Natur.458..993V} and the timescale inferred from near-earth asteroid (101955) Bennu scaled to main belt \citep{2020Sci...370.3660D}. Therefore, even the youngest dateable surface on Ceres could have undergone some degrees of weathering as opposed to being completely unweathered. As such, our understanding of the spectral evolution on Ceres is still very limited.

In this study, we focus on the spectral variation of the \SI{2.7}{\um} feature to provide insights into the possible space weathering process on Ceres. Specifically, we characterize the band center variation of the \SI{2.7}{\um} feature, its spatial distribution, and its correlation with the visible and near-infrared (VNIR) slope, albedo, and absorption strength. Combined with previous studies on surface age, we propose a new view of the spectral evolution on Ceres potentially associated with space weathering.

\section{Data and methods}

\subsection{Reflectance spectra}

The Dawn spacecraft mapped Ceres with its Framing Camera (FC) and Visible and Infrared Spectrometer (VIR), which acquired reflectance spectra from the visible to the near-infrared. The FC featured seven color filters to provide multispectral images in the \num{0.438}\text{--}\SI{0.965}{\um} range \citep{2011SSRv..163..263S}. We used photometrically corrected FC color data \citep{2014Icar..234...99S} to measure the VNIR slope (F5(\SI{965}{\nm})/F8(\SI{438}{\nm})) and the albedo variations. The imaging spectrometer VIR operates between \num{0.25} and \SI{5.0}{\um}, enabling the identification of a wide variety of minerals \citep{2011SSRv..163..329D}. For the analysis of spectral absorption features, we first processed the radiometrically calibrated VIR Level 1B data to remove the artifacts following \cite{2016RScI...87l4501C} and \cite{2020Life...11....9R}, and then subtracted the thermal emission longwards of $\sim$\SI{3.2}{\um} by fitting a blackbody spectrum with a temperature and emissivity to each spectrum \citep{2020Sci...370.3522S}. The thermally corrected radiance spectra were then divided by the distance-scaled solar irradiance to derive reflectance, and the photometric effect was corrected with the Hapke model using the parameters derived by \cite{2017A&A...598A.130C}. The final reflectance spectra were used to characterize the \SI{2.7}{\um}, \SI{3.1}{\um}, and \SI{3.9}{\um} absorption features attributed to phyllosilicates, ammoniated-clays \citep{2016Sci...353.4279A}, and carbonates \citep{2018SciA....4.1645C}, respectively.

We mainly used data acquired from the High Altitude Mapping Orbit (HAMO) phase with a spatial resolution of about \SI{450}{\meter\per\pixel} for VIR and \SI{140}{\meter\per\pixel} for FC. Due to the poor illumination conditions for the polar regions and the limited FC color data in the southern high latitude, we restricted our study to the latitude range between 50$^{\circ}$S and 60$^{\circ}$N. To minimize the effect of compositional variations in our analysis for space weathering effects, the known compositionally distinct regions were excluded, including the organic-rich Ernutet crater \citep{2017Sci...355..719D} and the Occator central region \citep{2020NatAs...4..786D}.

\subsection{Spectral analysis}

We characterized the \SI{2.7}{\um} absorption feature with two parameters: band center and band depth. Given that the spectral sampling interval of the VIR is about \SI{9.8}{\nm}, we performed a Gaussian fit to the absorption feature to improve the sensitivity to band center variation.

\begin{figure}[ht!]
\plotone{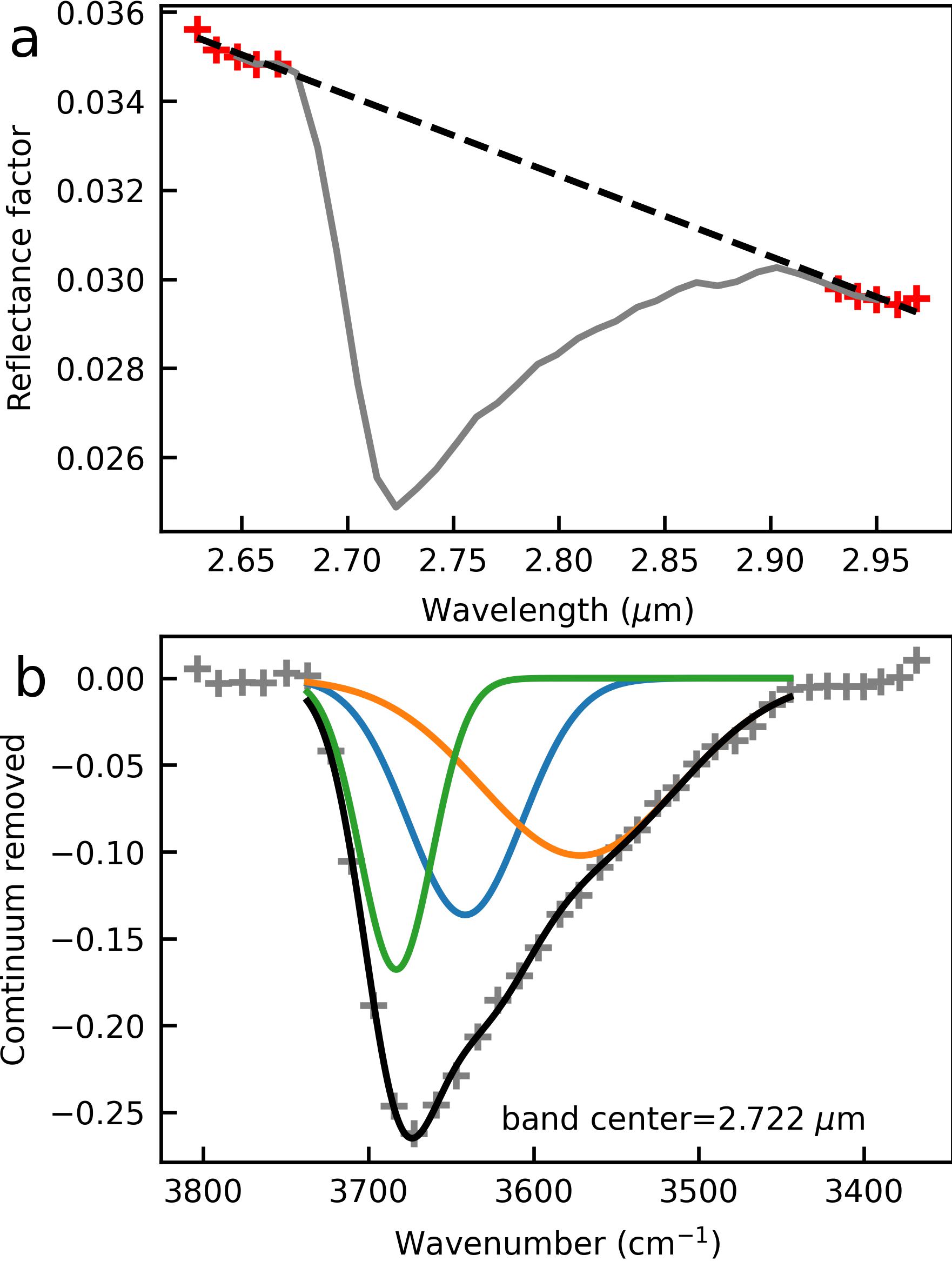}
\caption{\SI{2.7}{\um} absorption feature on Ceres. (a) An example of the \SI{2.7}{\um} absorption band. The black dashed line is the continuum calculated by a linear regression from the absorption shoulders marked with red crosses. (b) Three Gaussian functions fitted to the continuum removed \SI{2.7}{\um} absorption band. The band center derived from the model fitting is noted.    
\label{fig:spectral analysis}}
\end{figure}

The continuum of each spectrum was removed by a straight line (Fig. \ref{fig:spectral analysis}a), which is defined as a linear regression from both sides of the \SI{2.7}{\um} feature \citep{2023NatAs...7.1445L}. The anchor points are set at \num{2.63}\text{--}\SI{2.67}{\um} for the left shoulder and \num{2.93}\text{--}\SI{2.97}{\um} for the right shoulder. Then we used three Gaussian functions to model the \SI{2.7}{\um} absorption band in the wavenumber space to quantify band center and band depth (Fig. \ref{fig:spectral analysis}b), where band center is defined as the wavelength corresponding to the minimum of the best-fit continuum-removed spectrum, and band depth is defined as the ratio of the best-fit spectrum to the continuum at the band center. 

In order to validate the above procedure, we tested it with the laboratory spectra of similar materials with various spectral sampling, and compared the fitted band centers. The tests suggested that the band center can be retrieved at a precision higher than the spectral resolution of the data, as long as the signal-to-noise ratio is high enough, as for the case of the VIR data that we used. See Appendix \ref{sec:spectral resampling} for details.

Spectral ratio is also a commonly used method in spacecraft data analysis to highlight the spectral variations while minimizing the contribution of noise and artifacts \citep{2005Natur.438..623P, 2020A&A...644A.148S}. Here, we divided the spectrum with band center variation by the spectrum from the adjacent area based on the band center mapping results to verify the spectral variation.

\section{Results}

\subsection{\SI{2.7}{\um} band center variation}

\begin{figure}[ht!]
\plotone{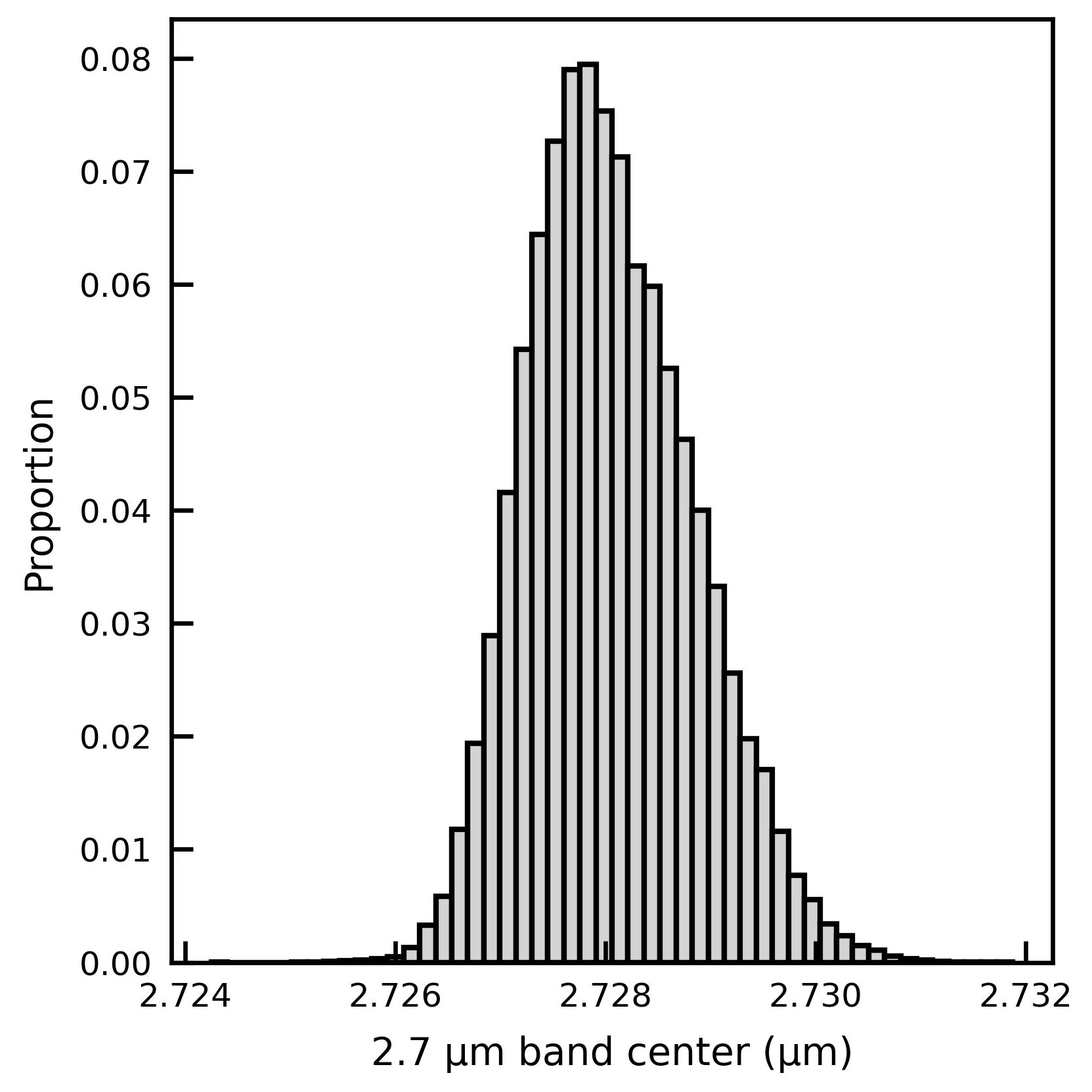}
\caption{Global distribution of the \SI{2.7}{\um} band center on Ceres as derived from the VIR HAMO data after spatially binned in 10$\times$10 pixels.
\label{fig:distribution of bc}}
\end{figure}

The \SI{2.7}{\um} band center is globally homogeneous, with values centered around $\sim$\SI{2.728}{\um} and primarily ranging from $\sim$\num{2.726} to
$\sim$\SI{2.730}{\um} (Fig. \ref{fig:distribution of bc}), consistent with the previous \SI{2.7}{\um} feature mapping results \citep{2016Sci...353.4279A}. However, some local regions exhibit band center shifts toward shorter wavelength by up to $\sim$\SI{5}{\nm} to $\sim$\num{2.723}\text{--}\SI{2.725}{\um} (hereafter referred to as blue-shifted regions) compared to adjacent regions. The distribution of the \SI{2.7}{\um} band center around the Haulani crater region is illustrated in Fig. \ref{fig:band center shift}a as an example. The other mapping results with blue-shifted regions are shown in Appendix \ref{sec:blue_shifted_spectra}. We note that these regions do not correspond to any shadows resolvable in the data. \added{We also examined the relationship between the \SI{2.7}{\um} band center and the slope of the linear continuum used for continuum removal (Fig. \ref{fig:band center shift}b). The two parameters show a negligible correlation (r $=$ \num{0.03}) with no systematic difference between the blue-shifted regions and background, suggesting that the observed band center shifts are unlikely to be introduced by the continuum removal procedure.} A comparison of the spectra in the blue-shifted and adjacent regions shows that the wavelength shift is not only for the band center, but for the entire \SI{2.7}{\um} absorption (Fig. \ref{fig:band center shift}c). To highlight such spectral variation, we ratioed the blue-shifted spectrum to the spectrum in the adjacent regions. The ratio spectrum shows a clear spike and dip feature (Fig. \ref{fig:band center shift}d), suggesting that the \SI{2.7}{\um} band center variations are real rather than artifacts potentially introduced by the Gaussian fit.

\begin{figure*}[ht!]
\plotone{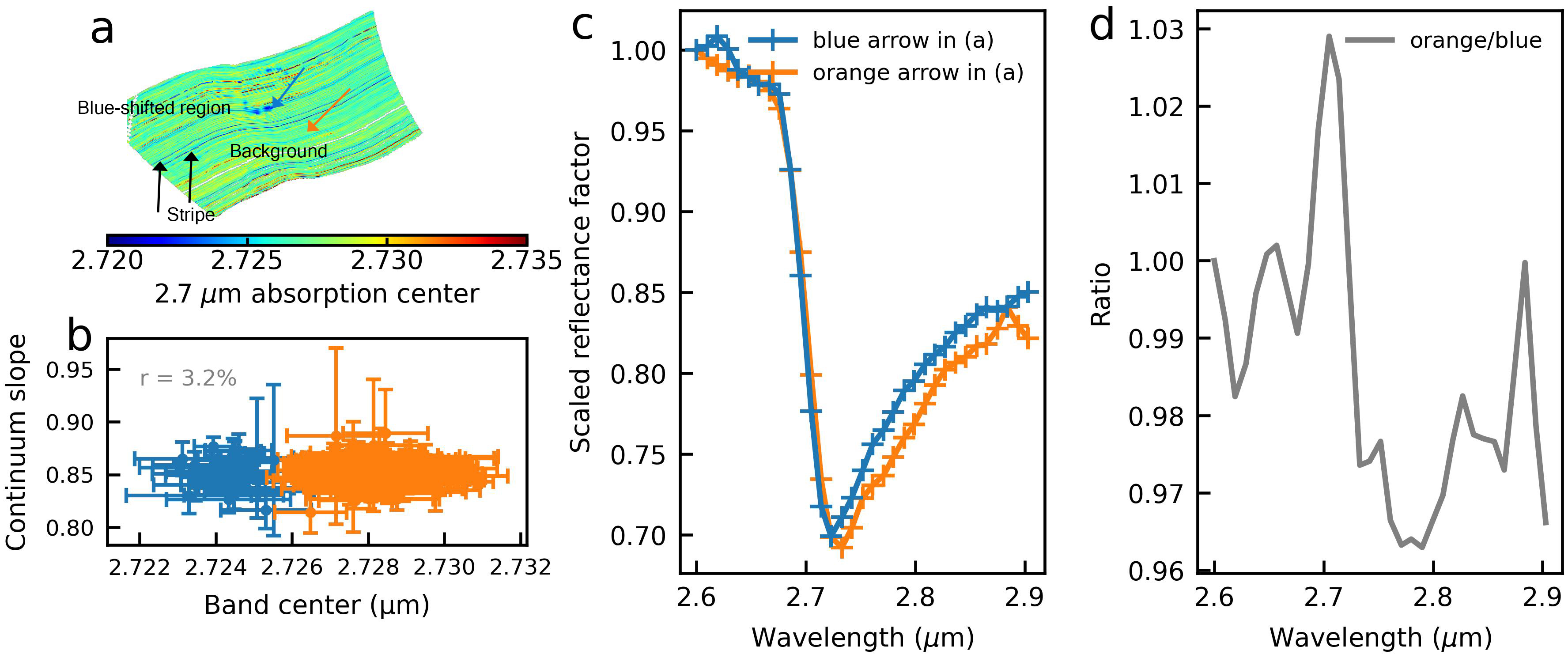}
\caption{Example of the \SI{2.7}{\um} band center variation. (a) \SI{2.7}{\um} band center map for the Haulani crater region. The blue and orange arrows indicate the locations of a blue-shifted region and an adjacent background region, respectively. \added{The black arrows indicate the stripe noise described in Section \ref{subsec:st1_distr}. (b) Correlation between the \SI{2.7}{\um} band center and the continuum slope of this band. The blue and orange points represent the blue-shifted regions and background, respectively.} (c) Scaled spectra of the blue-shifted and adjacent regions marked in (a). (d) Ratio of the background spectrum (orange) to the blue-shifted spectrum (blue) shown in (c).
\label{fig:band center shift}}
\end{figure*}

\subsection{Global distribution of blue-shifted regions}\label{subsec:st1_distr}

\begin{figure*}[ht!]
\plotone{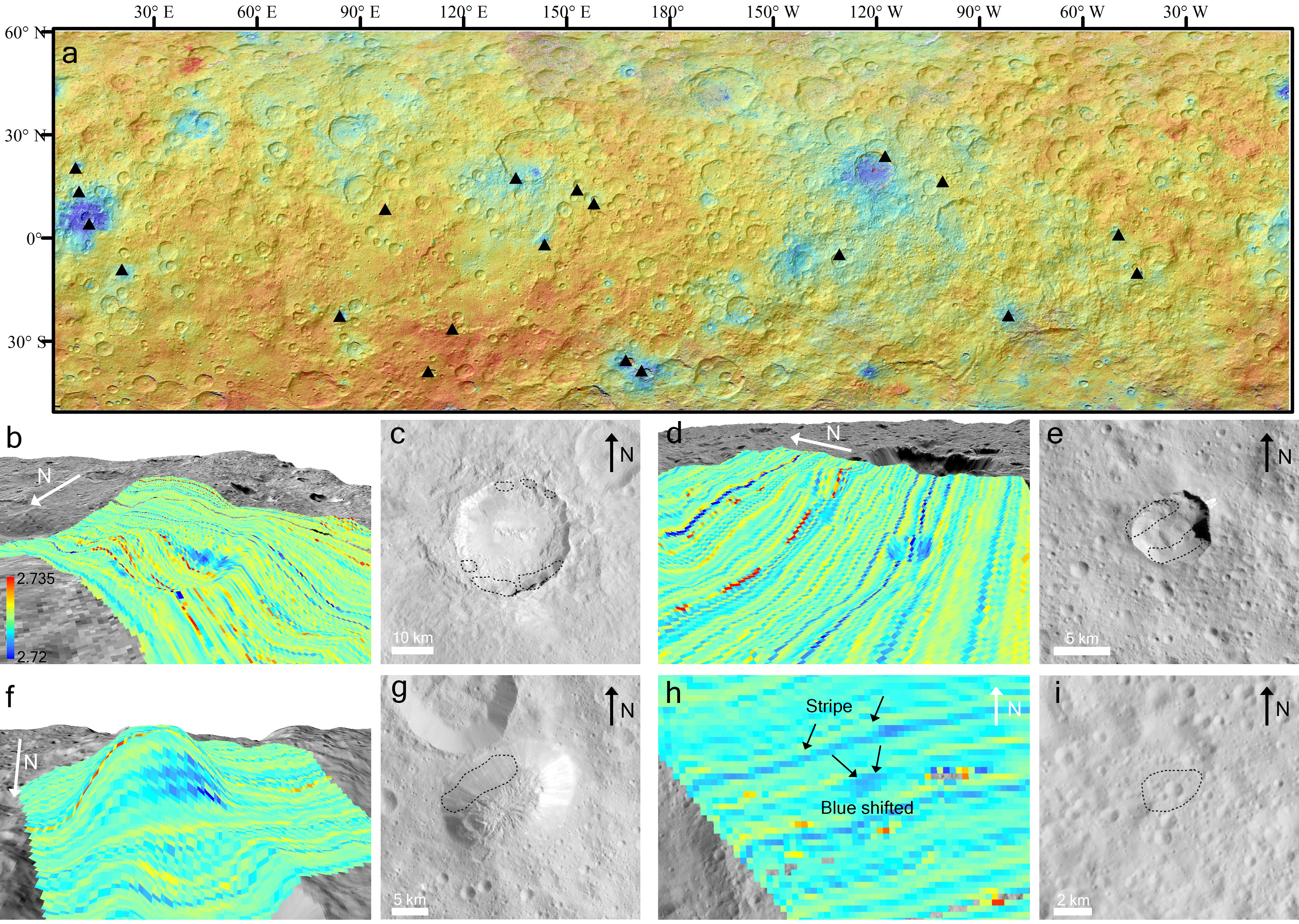}
\caption{Distribution of blue-shifted regions on Ceres. (a) The blue-shifted regions are marked with the black triangles. The background is color ratio F5(\SI{965}{nm})/F8(\SI{438}{nm}) superimposed on hillshade map. (b), (d) and (f) are 3d view of \SI{2.7}{\um} band center mapping around the Haulani crater, Braciaca crater and Ahuna Mons region, respectively. (c), (e) and (g) are FC clear filter image of the Haulani crater, Braciaca crater and Ahuna Mons region, respectively. The black dashed line illustrate the distribution of blue-shifted regions. (h) and (i) are \SI{2.7}{\um} band center mapping and FC clear filter image of one unnamed small crater region (100.64$^{\circ}$W, 16.43$^{\circ}$N), respectively. The black arrows in (h) illustrate the blue-shifted region and stripe noise. 
\label{fig:distribution}}
\end{figure*}

We conducted a global survey to identify blue-shifted regions over the whole surface of Ceres based on their \SI{2.7}{\um} band center. Blue-shifted regions are defined as those with their \SI{2.7}{\um} band center at wavelengths \textless\SI{2.726}{\um}. This boundary is selected based on the range of band centers of the overall Ceres surface, which is mostly between \num{2.726} and \SI{2.730}{\um}. Note that, due to the limitation of spatial resolution and data noise, some regions that exhibit blue-shifted band centers span only a few pixels and/or overlap with the stripe noise. Those regions are excluded from our analysis. Specifically, \added{stripe noise, which refers to spurious instrumental variations extending throughout an entire image column, was excluded by considering only those regions spanning multiple adjacent columns}. Whenever multiple observations were available for the same blue-shifted region, we verified the identification with different observations. An example of the identified blue-shifted region and the nearby stripe noise is provided in Fig. \ref{fig:distribution}h. 

Our global survey resulted in a total of \num{30} blue-shifted regions (Appendix \ref{table:blue_shifted}) \added{randomly} distributed across the surface of Ceres (Fig. \ref{fig:distribution}). \added{The mapping shows that the blue-shifted regions are distributed over a broad range of latitudes and longitudes, without any obvious geographic concentration}. Furthermore, comparison with the elemental abundance maps of H and Fe from Dawn's Gamma Ray and Neutron Detector (GRaND) \citep{PRETTYMAN201942}, the VNIR slope map (Fig. \ref{fig:distribution}a), and the phyllosilicates \citep{2016Sci...353.4279A} and carbonates \citep{2018SciA....4.1645C} maps \added{(Appendix \ref{sec:distribution_with_others}) suggests} that the distribution of blue-shifted regions is not correlated with any specific composition. 

On the other hand, the blue-shifted regions do show a correlation with topography. About \SI{75}{\%} of these regions are distributed on the crater walls, and their coverages are patchy rather than continuous (Figs. \ref{fig:distribution}b-e). One blue-shifted region is located on the north-western flank of Ahuna Mons (Figs. \ref{fig:distribution}f-g). Crater walls and mountain flanks generally have relatively steep slopes, which are susceptible to mass wasting. In fact, some of the regions with blue-shifted spectra have been interpreted as being associated with slumping materials, such as those in Haulani crater \citep{2016GeoRL..4311987S} and in Occator crater \citep{2019Icar..320...24N}. The downslope lineations on the Ahuna Mons flanks also indicate the presence of rock falls \citep{2016Sci...353.4286R}. These correlations suggest that the blue-shifted materials may be associated with mass wasting, which could expose relatively young subsurface materials. Additionally, the remaining \SI{25}{\%} blue-shifted regions are mainly associated with small craters of \textless\SI{2}{km} in diameter (Figs. \ref{fig:distribution}h-i), which generally represent relatively young surfaces as they are relatively more susceptible to being erased from the surface than large craters and thus must have formed more recently \citep{2024Icar..42016204Y}.

\subsection{Correlation between \SI{2.7}{\um} band center and other spectral features}

\begin{figure*}[ht!]
\plotone{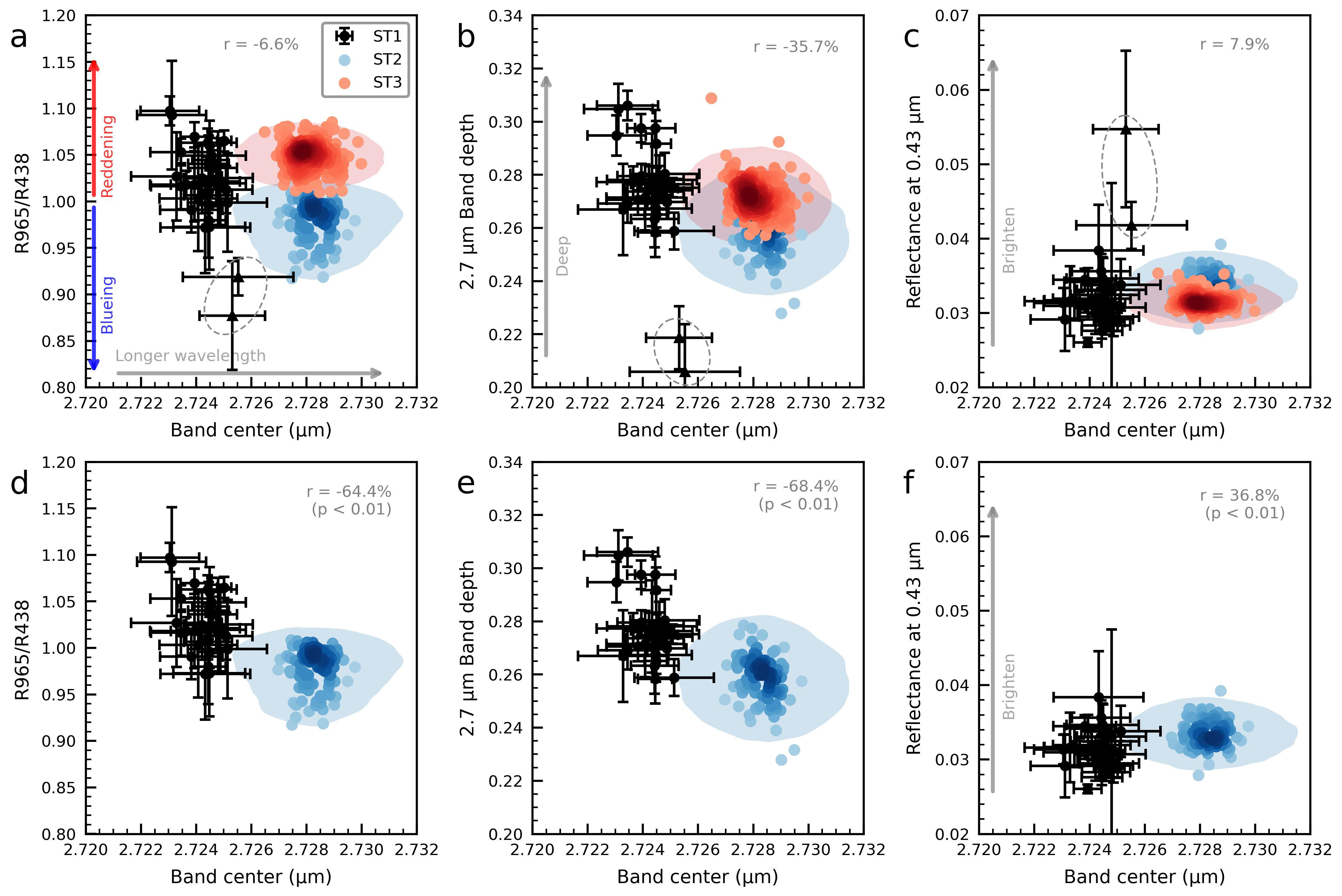}
\caption{Correlations between \SI{2.7}{\um} band center and other spectral features for all surface types (a-c) and for ST1 and ST2 only (d-f). The \SI{2.7}{\um} band center versus (a,d) \num{965}/\SI{438}{\nm} ratio; (b,e) \SI{2.7}{\um} band depth; (c,f) reflectance factor at \SI{438}{\nm}. The black, blue and red scatters represent the ST1, ST2 and ST3, respectively. The error bars for ST1 denote the standard deviations of the selected ROIs. For ST2 and ST3, color shading indicates the density of data points for those areas, and the semi-transparent colored regions mark the approximate ranges after accounting for the standard deviations of the corresponding spectral parameters. The correlation coefficients between \SI{2.7}{\um} band center and other spectral features were noted in the figure, excluding the two ST1 points enclosed by the gray oval, which may have different compositions.
\label{fig:spectral correlation}}
\end{figure*}

For a better understanding of the spectral difference between the blue-shifted regions and the rest of Ceres' surface, we analyzed their correlation with other spectral features: the VNIR spectral slope, the \SI{2.7}{\um} band depth, and albedo. Because previous studies suggested that the VNIR slope was a possible indicator of space weathering \citep{2016GeoRL..4311987S, 2017GeoRL..44.1660S, 2021NatCo..12..274S}, we chose the \SI{2.7}{\um} band center and VNIR slope to classify Ceres' surfaces into three types: Surface type 1 (ST1) represents the blue-shifted regions; \added{Surface type 2 (ST2) has a \SI{2.7}{\um} band center shifted toward longer wavelengths relative to ST1 and a \num{965}/\SI{438}{\nm} ratio less than 1; and Surface type 3 (ST3) has a similarly red-shifted \SI{2.7}{\um} band center relative to ST1, but a \num{965}/\SI{438}{\nm} ratio greater than 1}. The ST1 regions have been selected in the earlier analysis (Section \ref{subsec:st1_distr}). For ST2 and ST3, the presence of stripe noise in the VIR data and the lack of exact pixel-to-pixel correspondence between the VIR data and the FC data do not allow us to perform a pixel-wise statistic. Instead, for each VIR HAMO data cube in the entire dataset, we randomly selected regions of interest (ROIs) guided by the \SI{2.7}{\um} band center and the VNIR slope mappings. Each ROI should appear uniform in color and show no indications of shadows. This process will ensure a globally uniform distribution of the ST2 and ST3 regions without subjective bias. Finally, their spectral parameters were extracted by averaging within each ROI, and the corresponding standard deviations represent the spread of the parameters within each ROI (Fig. \ref{fig:spectral correlation}).

For all surface types, the \SI{2.7}{\um} band centers show no correlation with the VNIR slopes, the \SI{2.7}{\um} band depths, or the reflectances at \SI{438}{nm} (Figs. \ref{fig:spectral correlation}a-c). However, given that the different surface types may have different ages (see \S\ref{spectral_evolution}), we also analyze the correlations separately. For ST1 and ST2, which have relatively closer ages based on their spatial distributions, some correlations become evident. The \SI{2.7}{\um} band centers show moderate negative correlations (\num{-0.8} $<$ r $\le$ \num{-0.6}) with both the VNIR slopes (Fig. \ref{fig:spectral correlation}d) and the \SI{2.7}{\um} band depths (Fig. \ref{fig:spectral correlation}e), and a weak correlation (r $<$ \num{0.5}) with the reflectances at \SI{438}{\nm} (Fig. \ref{fig:spectral correlation}f). Between ST2 and ST3, although they have similar \SI{2.7}{\um} band centers, the spectrally red regions of ST3 exhibit deeper \SI{2.7}{\um} band depths and lower reflectances than the spectrally blue regions of ST2 (Figs. \ref{fig:spectral correlation}a-c). Such a correlation between ST2 and ST3 is consistent with previous studies \citep{2016GeoRL..4311987S, 2019Icar..318...56S}, demonstrating that our selection of the ROIs for these two types is representative of the whole surface of Ceres.

\section{Discussion}

\subsection{Cause of the \SI{2.7}{\um} band center shift \label{2.7micronshift}}

The wavelength shift of the \SI{2.7}{\um} band center and its correlation with relatively young surfaces on Ceres are intriguing, prompting us to investigate its implications for spectral evolution on Ceres. \added{The center wavelength of this band on Ceres could be affected by a number of factors, including: (1) surface physical properties; (2) mixing of other mineralogical compositions of Ceres \citep{2015Natur.528..241D}; (3) compositional variations in phyllosilicates due to various degrees of aqueous alteration \citep{2013MPS...48.1618T}; (4) mixing with exogenic materials \citep{2019NatAs...3..140M}; (5) space weathering \citep{2015AA...577A..41L}. In this section, we examine the above potential explanations for the \SI{2.7}{\um} band center shift on Ceres.}

\subsubsection{Surface physical properties}

\added{Variations in surface physical properties, such as grain size and porosity, can modify the optical path length and scattering behavior and thus affect the spectral slope, albedo, and band depth \citep{2021NatCo..12..274S, 2023ComEE...4..335M}. The \SI{2.7}{\um} band center is more directly related to the local bonding environment of the structural \ce{OH} groups and the crystal structure \citep{2008ClMin..43...35B}, and is less sensitive to these physical effects. In compositionally heterogeneous mixtures, however, differences in grain size among different mineral components could alter their relative optical contributions to the mixed spectrum and potentially affect the \SI{2.7}{\um} band center. Nevertheless, \cite{2023ComEE...4..335M} measured the Murchison meteorite powders with different grain sizes and porosities and reported that porosity variations do not significantly
affect the \SI{2.7}{\um} band center and that no \ce{OH}-band peak shift with grain size. We further quantified the potential effects of porosity using the laboratory spectra provided by \cite{2023ComEE...4..335M}. We processed those spectra using the same procedure applied to the VIR data, and derived the \SI{2.7}{\um} band center variations of less than \SI{2}{\nm} across different porosities. Therefore, we consider that the observed \SI{2.7}{\um} band center shift on Ceres is unlikely to be attributed to variations in porosity and grain size.}

\subsubsection{Mixing with other compositions}

Dawn mission observations have revealed that Ceres' surface is a complex mixture of Mg-phyllosilicates, ammoniated clays, carbonates, and dark components such as amorphous carbon or magnetite \citep{2015Natur.528..241D}. Some other compositions exist in a few localized areas, including hydrohalite (\ce{NaCl*2H2O}) and ammonium chloride (\ce{NH4Cl}) in Cerealia Facula inside Occator crater \citep{2020NatAs...4..786D}, \ce{H2O} ice \citep{2016Sci...353.3010C}, and organic materials \citep{2017Sci...355..719D, 2025AGUA....601362S} in a few regions. The dark materials are spectrally featureless around \SI{3}{\um} and therefore are not expected to affect the \SI{2.7}{\um} band. The ubiquitous \SI{2.7}{\um} absorption feature on Ceres is primarily attributed to the stretching vibration of \ce{O-H} coupled with a cation \citep{2008ClMin..43...35B}. However, the existence of other minor components in the mixture could also affect the exact wavelength and shape of this feature. For example, the $2\nu_3+\nu_4$ combination band of carbonates also occurs in the \num{2.7}\text{--}\SI{2.8}{\um} region \citep{2021E&SS....801844B}; therefore, variations in carbonate abundances could cause the center of the \SI{2.7}{\um} band to shift. Salts, such as the mixture of \ce{NH4Cl}, \ce{Na2CO3}, and \ce{NaCl*2H2O}, could also affect the \SI{2.7}{\um} band center due to the broad \SI{3.0}{\um} feature \citep{2019Icar..320..150T}.

\begin{figure}[ht!]
\plotone{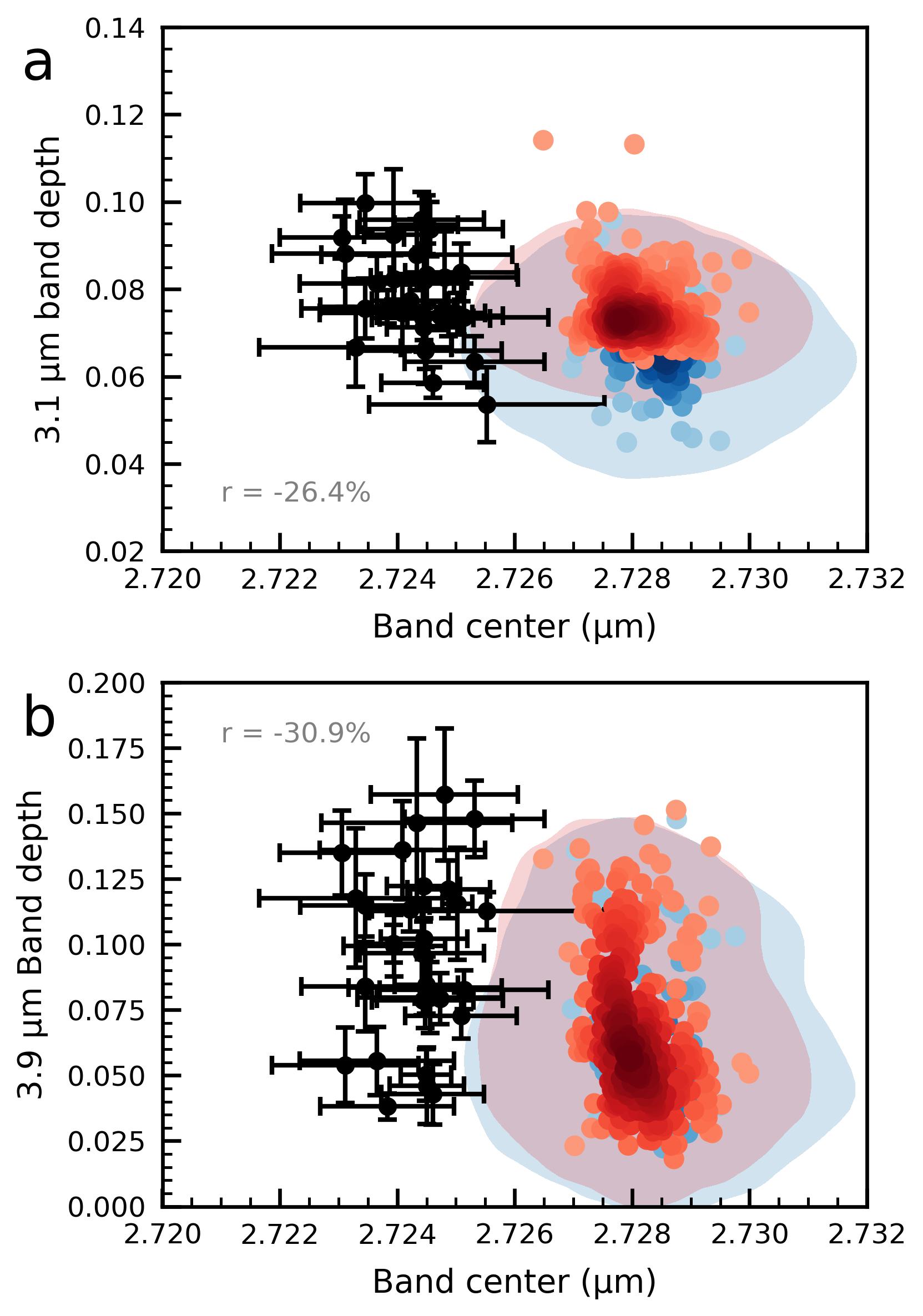}
\caption{Correlations between \SI{2.7}{\um} band center and the depths of \num{3.1} and \SI{3.9}{\um} bands across all surface types. The \SI{2.7}{\um} band center versus (a) \SI{3.1}{\um} band depth; (b) \SI{3.9}{\um} band depth. The black, blue and red scatters represent the ST1, ST2 and ST3, respectively. The \num{3.1} and \SI{3.9}{\um} band depths were calculated with the same Gaussian fit procedure. The error bars for ST1 denote the standard deviations of the selected ROIs. For ST2 and ST3, color shading indicates data density, and the semi-transparent colored regions mark the approximate ranges after accounting for the standard deviations. The correlation coefficients between \SI{2.7}{\um} band center and \num{3.1} and \SI{3.9}{\um} band depths were noted.
\label{fig:spectral correlation_bd}}
\end{figure}

To investigate whether the observed \SI{2.7}{\um} band center shift on Ceres could be caused by compositional variations, we performed two tests. First, we focused on carbonates and ammoniated clays using their diagnostic bands at \num{3.9} and \SI{3.1}{\um}, respectively. Previous studies have shown that the depths of those absorption features reflect variations in their abundances on Ceres \citep{2016Sci...353.4279A, 2018SciA....4.1645C}. Therefore, if the variations of their abundances caused the \SI{2.7}{\um} band center shift, then the band center should show a correlation to some extent with the depth of their diagnostic bands. However, \added{the correlation coefficients of approximately \num{-0.2}\text{--}\num{-0.3} suggests a weak to no correlation} (Fig. \ref{fig:spectral correlation_bd}). We note that these correlation analyzes were performed across all surface types, because a compositional control on the \SI{2.7}{\um} band center would be expected to be across the global surface rather than be confined to specific surface types. These results suggested that the \SI{2.7}{\um} band center shift should not be dominated by the abundance variations of carbonates or ammoniated clays mixed in the regolith of Ceres.

For the second test, we aimed to quantify the sensitivity of the \SI{2.7}{\um} band center to the abundance variations of \ce{H2O} ice and salts. Although these components have been identified only in a few localized regions on Ceres \citep[e.g.,][]{2016Sci...353.3010C, 2019Icar..318...22C, 2020NatAs...4..786D}, their potential presence at undetectable levels of abundance in other areas cannot be ruled out. Here, we used a linear spectral mixing model to add small amounts of water ice and salts, respectively, to the average spectrum of Ceres and quantified how much band center shift we would detect (Appendix \ref{sec:sensitivity analysis}). Our analysis showed that minor amounts of \ce{H2O} ice or salts with abundances undetectable in the spectrum \added{(relative abundances of less than \SI{0.5}{\%})} would cause \textless\SI{1}{nm} shift for the \SI{2.7}{\um} band center, comparable to the uncertainty of the band center shift that we detected.

With these two tests, we concluded that minor amounts of carbonates, salts, or water ice mixed in Ceres regolith are unlikely to drive the observed variations in the \SI{2.7}{\um} band center. Other minor minerals, especially in the hydrated phases without other spectral features in the VIR spectral range, could also affect the exact wavelength position of the \SI{2.7}{\um} band. Until more high-resolution data or samples are available in the future, we cannot prove or rule out such a possibility. However, we suggest that the distribution of those minor phases, if excavated from the subsurface, should be widely distributed on the crater floor, wall, and ejecta \citep{1989icgp.book.....M}, a pattern widely observed on other planetary bodies \citep[e.g.][etc.]{2015Icar..248..373C,2024PSJ.....5..114A}. The fact that the blue-shifted regions only occur on the crater walls does not seem to support this scenario.

\subsubsection{Compositional variations of phyllosilicates}

In phyllosilicates, the \ce{OH} stretching band center varies with the Mg/Fe value. Substitution of Mg by Fe causes the band center to shift toward longer wavelengths \citep{2008ClMin..43...35B}. Different degrees of aqueous alteration change the Mg/Fe value in phyllosilicates. The most altered carbonaceous chondrites exhibit a band center at \num{2.71}\text{--}\SI{2.72}{\um}, which is consistent with antigorite (Mg-rich serpentine), whereas the least altered carbonaceous chondrite is rich in cronstedtite (Fe-rich serpentine), characterized by the band center at $\sim$\SI{2.80}{\um} \citep{2013MPS...48.1618T}. The identified blue-shifted regions are mainly associated with crater walls, representing impact excavation materials. If the observed band center variations are attributed to different degrees of aqueous alteration between surface and subsurface materials, as discussed in the previous section, a similar wavelength shift should also occur in the crater floor and ejecta regions, \added{where there are other} excavated subsurface materials. However, we did not observe such a trend.

Additionally, in the Ahuna Mons region, which is accepted as a cryovolcanic dome \citep{2016Sci...353.4286R} and could present a different phyllosilicate composition, the blue-shifted regions are mainly distributed on the north-western flanks rather than \added{throughout the flanks and summit. If the band center shift were primarily controlled by the composition of the cryomagmatic material, a more widespread correspondence with the dome would be expected. The localized distribution therefore suggests that cryomagmatic composition is unlikely to be the primary control on the \SI{2.7}{\um} band center variation.}

\subsubsection{Exogenic materials}

The contamination from exogenic materials may also influence the \SI{2.7}{\um} band, as has been found almost everywhere on Vesta \citep{2012ApJ...758L..36D, 2012Natur.491...83M, 2012Icar..221..544R}. \cite{2019NatAs...3..140M} suggested that Ceres' surface is best explained by the presence of \num{50}\text{--}\SI{60}{\%} CC-like materials, possibly due to infalling materials from asteroid impacts. The admixture of endogenic and exogenic carbonaceous materials could contribute to variations in the phyllosilicate bands. However, this scenario remains difficult to reconcile with the spatial distribution of the blue-shifted regions identified in this study. On Vesta, the exogenic carbonaceous materials are often associated with craters, appearing mostly in ejecta and crater walls \citep{2012Natur.491...83M}. In contrast, our identified blue-shifted regions are predominantly confined to the walls of some craters but not observed in the corresponding ejecta. This discrepancy suggests that the \added{materials associated with the observed \qty{2.7}{\um} band center variations seem to have a different emplacement mechanism than that of the exogenic carbonaceous materials on Ceres.}

\subsubsection{Space weathering}

Several laboratory experiments with carbonaceous chondrites and phyllosilicates showed that space weathering could induce a shift of the \SI{2.7}{\um} band center toward longer wavelengths \citep{2015AA...577A..41L, 2017Icar..285...43L, 2020PSJ.....1...61R}. \added{Remote-sensing observations of Bennu revealed a shortward shift of the \SI{2.7}{\um} band center in relatively fresh material exposed within craters compared with the global average spectrum \citep{2021Icar..35714252D}.} The recently returned samples from Ryugu reveal that the \SI{2.7}{\um} band center of the unweathered materials collected from the subsurface excavated by the artificial impact is shorter than that of the presumably weathered surface samples by $\sim$\SI{6}{\nm} \citep{2023Icar..40615755H, 2023NatAs...7.1445L}, which is comparable to what we observed on Ceres. In addition, the correlation between the blue-shifted regions on Ceres and relatively young surface materials (Fig. \ref{fig:distribution}) suggests that surface exposure age variations might be responsible for the wavelength shift of the \SI{2.7}{\um} band. Therefore, we consider space weathering as a possible explanation for the observed \SI{2.7}{\um} band center shift on Ceres.

However, we must recognize that the space weathering process in the main asteroid belt could be very different from that in near-Earth space, where we gain most of our knowledge about this process. Although traditional space weathering such as the production of nanophase or microphase opaques could still exist in the main belt, other processes, such as accumulation of exogenous materials \citep[e.g.,][]{2019NatAs...3..140M, 2012Natur.491...83M}, regolith mixing \citep[e.g.,][]{2012Natur.491...79P}, and perhaps some yet to be identified processes, could all contribute to the regolith evolution and spectral change on the main-belt asteroids. In this regard, although we tried to be comprehensive, our work will inevitably be limited in the interpretations of the spectral evolution on Ceres. Therefore, we limited the scope of this work to identifying the possible spectral evolutionary trend on Ceres and discussing the possible explanations in terms of our traditional understanding of space weathering.

\subsection{Spectral evolution on Ceres}\label{spectral_evolution}

Under the assumption that the small \SI{2.7}{\um} band center shift on Ceres represents a spectral evolution, we can constrain the timescale of such process from the ages of the associated geological features. The association of blue-shifted regions with mass wasting on the crater walls indicates that those regions must be younger than their host craters. If we assume all blue-shifted regions undergo similar spectral evolution, then their ages must be younger than the youngest host craters. The youngest dateable crater associated with the blue-shifted regions, the Cacaguat crater, has a Lunar-derived model (LDM) age of $\sim$1.3 Ma \citep{2018Icar..316...99W}. Thus, this age places an upper limit of $\sim$\num{e6} years for the process responsible for the \SI{2.7}{\um} band center shift. In order to explain why not all mass wasting regions host areas with blue-shifted \SI{2.7}{\um} band, we suggest that those mass wasting may have occurred before \num{e6} years ago and already been weathered. It has to be recognized, however, that it is impossible to estimate the exact ages of the blue-shifted regions, and of all other mass wasting areas that do not show blue-shifted \SI{2.7}{\um} band center. This hypothesis can be tested when high-resolution data become available from future missions to Ceres. Because the timescale of the \SI{2.7}{\um} band center shift is shorter than that of the previously reported spectral color variations on Ceres \citep{2016GeoRL..4311987S, 2017GeoRL..44.1660S}, we suggest that the \SI{2.7}{\um} band center shift may represent an early-stage evolution that occurs in a time frame of \textless\num{e6} years. \added{The blue-shifted ST1 regions represent the youngest or most recently exposed materials. With their exposure ages increasing, their \SI{2.7}{\um} band centers are expected to shift toward longer wavelengths and eventually become ST2-like.}

The spectral characteristics of the early-stage evolution are illustrated by the correlations of the spectral features of the ST1 and ST2 regions. The moderate negative correlations between the \SI{2.7}{\um} band center and the VNIR slope, as well as with the \SI{2.7}{\um} band depth (Figs. \ref{fig:spectral correlation}d-e), indicate a trend with spectrally blueing and weakening of the \SI{2.7}{\um} band. If we assume that the \SI{2.7}{\um} band center evolution is due to space weathering, as discussed in \S\ref{2.7micronshift}, then such correlations suggest that space weathering could be a dominant contributor to the VNIR slope and the \SI{2.7}{\um} band depth variations. However, other factors, such as the phyllosilicate abundances in the mixture and surface physical properties, could also affect those spectral features. The relatively weak correlation between the \SI{2.7}{\um} band center and the reflectance at \SI{438}{\nm} (Fig. \ref{fig:spectral correlation}f) suggests that space weathering may not be the controlling factor for albedo variation, but its effect cannot be entirely ruled out.

\begin{figure*}[ht!]
\plotone{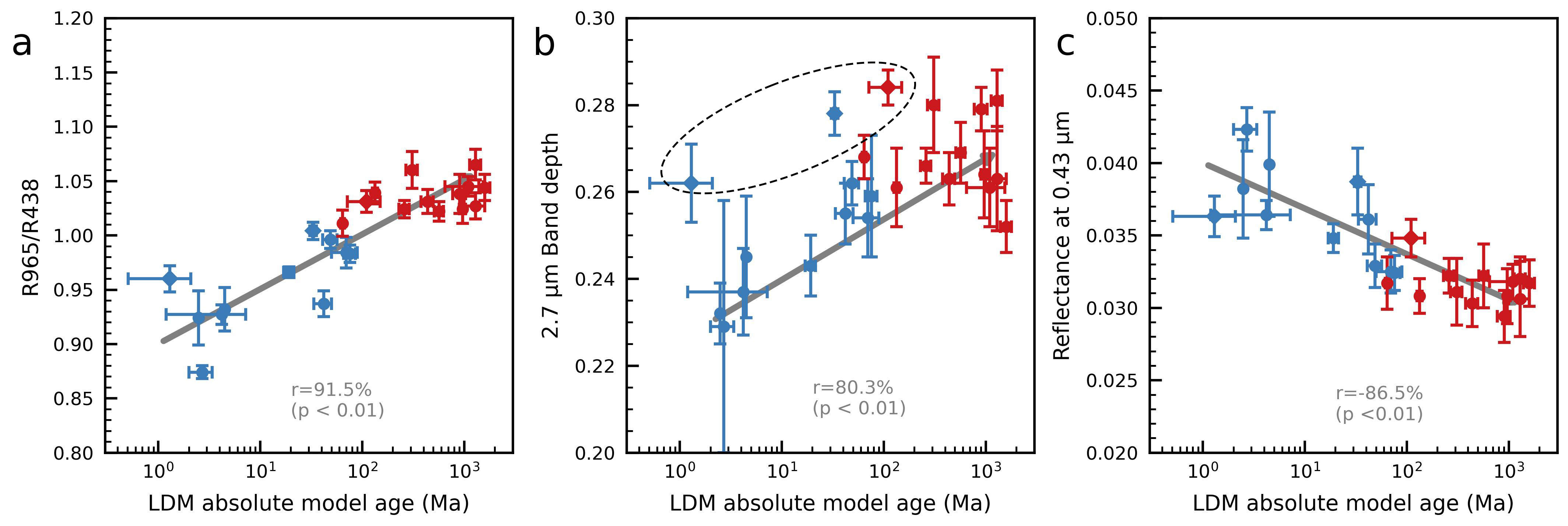}
\caption{Correlations between the LDM and the corresponding spectral features of some typical regions \citep{2016GeoRL..4311987S, 2019Icar..318...56S}. The LDM verses (a) \num{965}/\SI{438}{\nm} ratio; (b) \SI{2.7}{\um} band depth; (c) reflectance factor at \SI{438}{\nm}. The blue and red scatters represent the ST2 and ST3, respectively. The gray oval illustrates the outliers in \SI{2.7}{\um} caused by compositional differences \citep{2019Icar..318...56S}: Dantu, Rao and Cacaguat crater.
\label{fig:correlation with age}}
\end{figure*}

The previous studies about the spectral reddening on Ceres with time provide clues about the spectral evolution on a timescale longer than $\sim$\num{e6} years \citep{2016GeoRL..4311987S, 2019Icar..318...56S}, which we refer to as late-stage spectral evolution here. Our work shows that the \SI{2.7}{\um} band center stops shifting in this stage, centered around \SI{2.728}{\um}, but the spectral color on Ceres continues to evolve as suggested by the trend with the ages of craters (Appendix \ref{sec:age}). Specifically, both the VNIR spectral slope and the \SI{2.7}{\um} band depth show a strong positive correlation with surface age \added{as determined from crater counting} (Figs. \ref{fig:correlation with age}a-b), whereas the reflectance at \SI{438}{\nm} has a strong negative correlation with surface age (Fig. \ref{fig:correlation with age}c). These age-dependent correlations suggest that the observed spectral evolution could be attributed to space weathering. However, albedo, spectral slope, and absorption strength are also sensitive to surface composition and physical properties \citep{2016Sci...353.4279A, 2023ComEE...4..335M}, potentially contributing to the observed spectral variations. Indeed, a few outliers were identified that likely result from the surface composition variations \citep{2019Icar..318...56S} (Fig. \ref{fig:correlation with age}b). Overall, late-stage spectral evolution on Ceres is characterized by lowering the albedo, reddening the VNIR slope, and strengthening the \SI{2.7}{\um} feature.

Putting both stages together, it appears that the spectral evolution of Ceres' surface is not monotonic. \added{In our proposed evolutionary sequence, the blue-shifted ST1 regions represent the youngest materials and correspond to relatively fresh subsurface material underlying the more evolved ST2 and ST3 (Table \ref{table:1}). During the early-stage, as exemplified by the evolution from ST1 to ST2, the freshly exposed materials become spectrally bluer, the \SI{2.7}{\um} band weakens, and its center shifts toward longer wavelengths within about a million years. The trend then reverses during the late-stage, as demonstrated by the comparison between ST2 and ST3: the VNIR slope reddens and the \SI{2.7}{\um} band deepens with increasing exposure age, while the band center remains relatively unchanged. Because the VNIR slope evolves in opposite directions during the two stages, similar VNIR slopes may occur at different exposure ages. Consequently, the VNIR slope alone cannot uniquely identify the youngest blue-shifted regions or distinguish them from more evolved surfaces.}

\begin{deluxetable*}{ccccc}
\tablewidth{0pt}
\tablecaption{Summary of spectral features at different stages \label{table:1}}
\tablehead{
\colhead{Surface type} & \colhead{\SI{2.7}{\um} BC}\tablenotemark{a} & \colhead{VNIR slope} & \colhead{\SI{2.7}{\um} BD}\tablenotemark{b} & \colhead{Age}
}
\startdata
ST1 & Unshifted & Red & Deep & Young \\ 
ST2 & Red shifted & Blue & Shallow & Intermediate \\
ST3 & Red shifted & Red & Deep & Old \\
\enddata                          
\tablenotetext{a}{BC: band center}
\tablenotetext{b}{BD: band depth}
\end{deluxetable*}

\subsection{Mechanisms for spectral evolution}

In this section, we discuss possible mechanisms that could cause the spectral behavior we observed on Ceres. However, we emphasize that it is premature to definitively determine the mechanisms with our current understanding of the spectral evolution on carbonaceous materials. Further laboratory experiments targeting the specific composition of Ceres, or even a sample return, would be necessary.

For the \SI{2.7}{\um} band center, previous discussion (\S\ref{2.7micronshift}) suggests that the observed shift may be associated with space weathering. Specifically, laboratory studies have revealed that ion implantation (simulation of solar wind irradiation) can cause the \SI{2.7}{\um} band center to shift toward longer wavelengths \citep{2015AA...577A..41L, 2017Icar..285...43L}, while laser irradiation (simulation of micrometeorite impact) does not have such an effect \citep{2020ApJ...890L..23M, 2022Icar..37214736P}. Therefore, we suggest that the solar wind bombardment may play a key role in modifying the \SI{2.7}{\um} phyllosilicate feature during the early-stage spectral evolution. Ion bombardment may preferentially sputter the relatively light Mg, and/or preferentially cause amorphization of Mg-rich materials, leading to a decrease of the Mg/Fe ratio in phyllosilicates \citep{2014Icar..237..278B, 2015AA...577A..41L}, and therefore an increase of the wavelength of the \SI{2.7}{\um} band center.

Regarding the changes in spectral slope, albedo, and band depth on Ceres, many factors can contribute, including chemical/mineralogical composition, physical property, or both.

First, we consider \added{the possibility} that the spectral evolution on Ceres \added{is} attributed to chemical/mineralogical factors, including space weathering products and/or exogenic materials. Space weathering commonly produces metallic iron, magnetite, troilite, and graphitized carbon in primitive materials \citep{2015LPI....46.1913K, 2015Icar..254..135M, 2020Icar..34613775T, 2021Icar..36414479L, 2025NatGe..18..825K}. Laboratory studies and theoretical modeling have shown that nano-phase Fe and troilite can cause spectral reddening, while graphite and magnetite cause spectral blueing \citep{2020Icar..34613775T, 2021PSJ.....2...68T, 2021M&PS...56.1173S}. However, even if the spectral color changes on Ceres could potentially be explained by perhaps different preferential production of different space weathering products in different stages, a multi-stage weathering process was rarely observed in samples or meteorites, although not entirely impossible \citep{2017Icar..292..245K, 2022A&A...659A..78Z}. \added{Also, both weakening \citep{2015AA...577A..41L} and strengthening \citep{2021Icar..36414479L, 2021Icar..35514140N} of the \SI{2.7}{\um} feature have been observed in laboratory experiments that simulate space weathering. The band weakening may result from the amorphization and dehydration of the phyllosilicates, whereas \ce{H+} irradiation can also produce \ce{H2O} and hydroxyl species, potentially strengthening the hydration band. However, no experiment to date has shown a process with band depth decreases first and then increases.} Given that Ceres' composition is distinct from all meteorites in the current collections \citep{2018M&PS...53.1793M}, we have no base to explain the possible space weathering products on it.

Alternatively, exogenic materials could also contribute to the observed spectral evolution on Ceres. \cite{2019NatAs...3..140M} suggested that \added{Ceres'} surface is a mixture of endogenic materials and \num{50}\text{--}\SI{60}{\%} exogenic CC-like materials. In this context, the increase in the \SI{2.7}{\um} band depth in the late-stage may be explained by the progressive, secular accumulation of CC-like material on the surfaces. This interpretation, involving changes in the relative abundances of compositional components, may also account for other age-related spectral trends in the late stage. However, the \SI{2.7}{\um} absorption band of phyllosilicates is significantly stronger than that of CC-like materials \citep{2020JGRE..12506606K}. The accumulation of CCs would dilute the relative abundance of phyllosilicates in the mixture and consequently weaken the \SI{2.7}{\um} absorption band. Therefore, the accumulation of CC-like materials is not likely the reason for the observed increase in band strength with age.

Changes in physical properties induced by space weathering would also affect spectral evolution \citep{2016JGRE..121.1865P}. The early-stage spectral evolution could be explained by an increase in grain size, as coarser grains generally exhibit spectrally bluer slopes and weaker hydration features \citep{2011Icar..212..180C}. But the mechanism for aggregating grains is poorly understood, and it is difficult to explain why the albedo does not show an appreciable trend as it is also affected by grain size. Previous work has attributed the spectral reddening during the late-stage to a decrease in grain size \citep{2017GeoRL..44.1660S} and/or porosity \citep{2021NatCo..12..274S}. \added{Such changes in surface physical properties could potentially account for the observed late-stage band strengthening.} However, a decrease in grain size and/or porosity should also increase albedo \citep{2023ComEE...4..335M}, a phenomenon that contradicts observations (Fig. \ref{fig:correlation with age}c). Therefore, if changes in physical properties cause the spectral variations on Ceres, then multiple and potentially competing factors may dominate different stages.

\subsection{Comparison with other C-complex asteroids}

To put the spectral evolution that we postulate for Ceres into a broader context, here we compare it with the spectral trend of other C-complex asteroids.

Telescopic surveys of C-complex asteroid families suggest that the spectral changes in primitive-type asteroids, presumably induced by space weathering, are diverse and complex. \cite{2005Icar..173..132N} studied color variations between young and old C-complex asteroid families using photometric data and found that they become spectrally blue over time. \cite{2016Icar..264...62K} investigated the Beagle and Themis families, which were thought to originate from the same parent body and therefore have the same composition but different ages. Their results, however, showed that space weathering leads to reddening and darkening. \cite{2006ApJ...647L.179L}, using a smaller set of spectroscopic data, also found a reddening trend for C-complex asteroids. Interestingly, more recently, \cite{2021AJ....161...99T} performed a study of spectral slope as a function of size for nine C-complex families and assumed that smaller objects have younger surfaces on average. They found two distinct spectral slope trends. One trend, termed Themis-type, showed a spectral reddening with increasing object size until a maximum slope value was reached. The other, Hygiea-type families, showed a spectral blueing with increasing object size until a minimum slope value was reached, and then reddening with increasing object size, a spectral trend similar to what we suggested for Ceres. All relevant studies emphasized that the complex compositional variations in C-complex asteroids may be the reason for their diverse spectral behaviors of space weathering.

In situ observations of Ryugu and Bennu also revealed different space weathering effects. \cite{2024Icar..42016204Y} suggested that the fresh surfaces of Ryugu and Bennu initially had similar spectra, but subsequently evolved in opposite directions due to space weathering. The surface of Ryugu became redder and darker \citep{2020Sci...368..654M}, accompanied by a \SI{2.7}{\um} band center shift and band weakening, similar to the early-stage spectral evolution that we suggested for Ceres \citep{2023Icar..40615755H, 2023NatAs...7.1445L}. On the other hand, space weathering on Bennu exhibits a non-monotonic spectral behavior similar to that of Hygiea-type families revealed by telescopic observations. Fresh materials on Bennu first became bluer and then redden over time, possibly by brightening in the blue band first and then in the red later \citep{2020Sci...370.3660D, 2023Icar..40015563C}. This trend is consistent with our suggested spectral trend of Ceres' surface. In terms of the overall spectral slope, Bennu appears to be blue, and is bluer than that of Ceres, which also has a slightly blue spectrum at low phase angles \citep{2016ApJ...817L..22L}, although the implications are unclear. Furthermore, both the \SI{2.7}{\um} band center reddening and the \SI{2.7}{\um} band weakening also occur on Bennu, while, unlike Ceres, the band depth trends are monotonic \citep{2020Sci...370.3660D, 2023Icar..40015563C}.

All those studies suggest that: 1) The spectral behaviors of space weathering on C-complex asteroids are complex and diverse, likely due to their large compositional variations compared to, e.g., the S-complex asteroids. 2) Multistage spectral evolution may be common on carbonaceous asteroids, complicating the interpretations of the surface composition and evolution of C-complex asteroids from spectral observations. And 3) The spectral slope trend that we suggest for Ceres has been observed elsewhere. In fact, it is consistent with what was observed for the Hygiea family. Note that Hygiea shares similar spectral features with Ceres \citep{2025PSJ.....6....9R}, indicating a similar surface composition and thereby possibly similar spectral evolution processes.

\section{Conclusions} \label{sec:cite}
Our findings provide new insights into the spectral evolution on Ceres, potentially associated with space weathering. The spectral analysis of the Dawn multispectral imaging and spectroscopic data reveals that the spectra of some crater wall materials and small craters have the center of the \SI{2.7}{\um} band shifted to shorter wavelengths compared to those of most of Ceres' surface. Our analysis suggests that space weathering is the most plausible explanation for the band center shift. The youngest crater associated with the band center shift is about 1 Ma, putting a timescale for the band center shift. Therefore, the \SI{2.7}{\um} band center can be used as an indicator of surface maturation within a timescale of $<$ \num{e6} years.

Combined with the spectral reddening of Ceres on a longer timescale as reported in the previous studies and interpreted as a possible space weathering effect, our observations suggest that the spectral evolution of Ceres' surface presents a two-stage rather than a monotonic process. Freshly exposed materials on Ceres initially become bluer with the \SI{2.7}{\um} band strength weakening in a timescale of about a million years, together with \SI{2.7}{\um} band center shifting to longer wavelengths by a few \si{nm}. The trend then reverses with spectral reddening and strengthening of the \SI{2.7}{\um} band in a longer time frame, but the \SI{2.7}{\um} band center remains unchanged. These spectral variations may be attributed to the chemical/mineralogical and/or physical changes associated with space weathering. Such behavior is broadly consistent with the spectral slope trend of space weathering on Bennu and the Hygiea family asteroids. Our study provides insights into the evolution of Ceres' regolith and the understanding of the diverse spectral evolution on C-complex asteroids.

\begin{acknowledgments}

J.-Y. Li acknowledges the support by the 2024 Xinjiang Autonomous Region Tianchi Talent Program and by Natural Science Foundation of Xinjiang Uygur Autonomous Region No. 2025D01E62. Y.Z. Yang acknowledges the support from the Beijing Nova Program. \added{The authors would like to thank Prof. Beth Ellen Clark and one anonymous reviewer for their constructive reviews.} We thank the VIR team for providing the calibration data and Dr. Mauro Ciarniello for sharing the photometric parameters. We thank Dr. Tania Le Pivert-Jolivet for insightful discussions and are grateful to Dr. Maria Cristina De Sanctis for her constructive suggestions on this work. The authors thank the Dawn mission team for the development, cruise, orbital insertion, and operations of the Dawn spacecraft at Ceres. Dawn data are archived in NASA's Planetary Data System Small Bodies Node. The VIR data is available at \url{https://arcnav.psi.edu/urn:nasa:pds:context:instrument:dawn.vir} and the FC data is available at \url{https://arcnav.psi.edu/urn:nasa:pds:context:instrument:dawn.fc2}. \added{The derived maps are available at \url{https://arcnav.psi.edu/urn:nasa:pds:context:investigation:mission.dawn_mission_to_vesta_and_ceres/more}}

\end{acknowledgments}

\begin{contribution}

Q. Zhang led the conceptualization, data curation and analysis, interpretation, and writing of the paper. J.-Y. Li contributed to the project supervision, conceptualization, data analysis, and reviewed and edited the paper. Y. Yang, M. Liu and Y. Liu contributed to the interpretation, and reviewed and edited the paper.

\end{contribution}





\appendix

\section{The impact of VIR spectral sampling on the \SI{2.7}{\um} band center fit} \label{sec:spectral resampling}

\begin{figure}[ht!]
\plotone{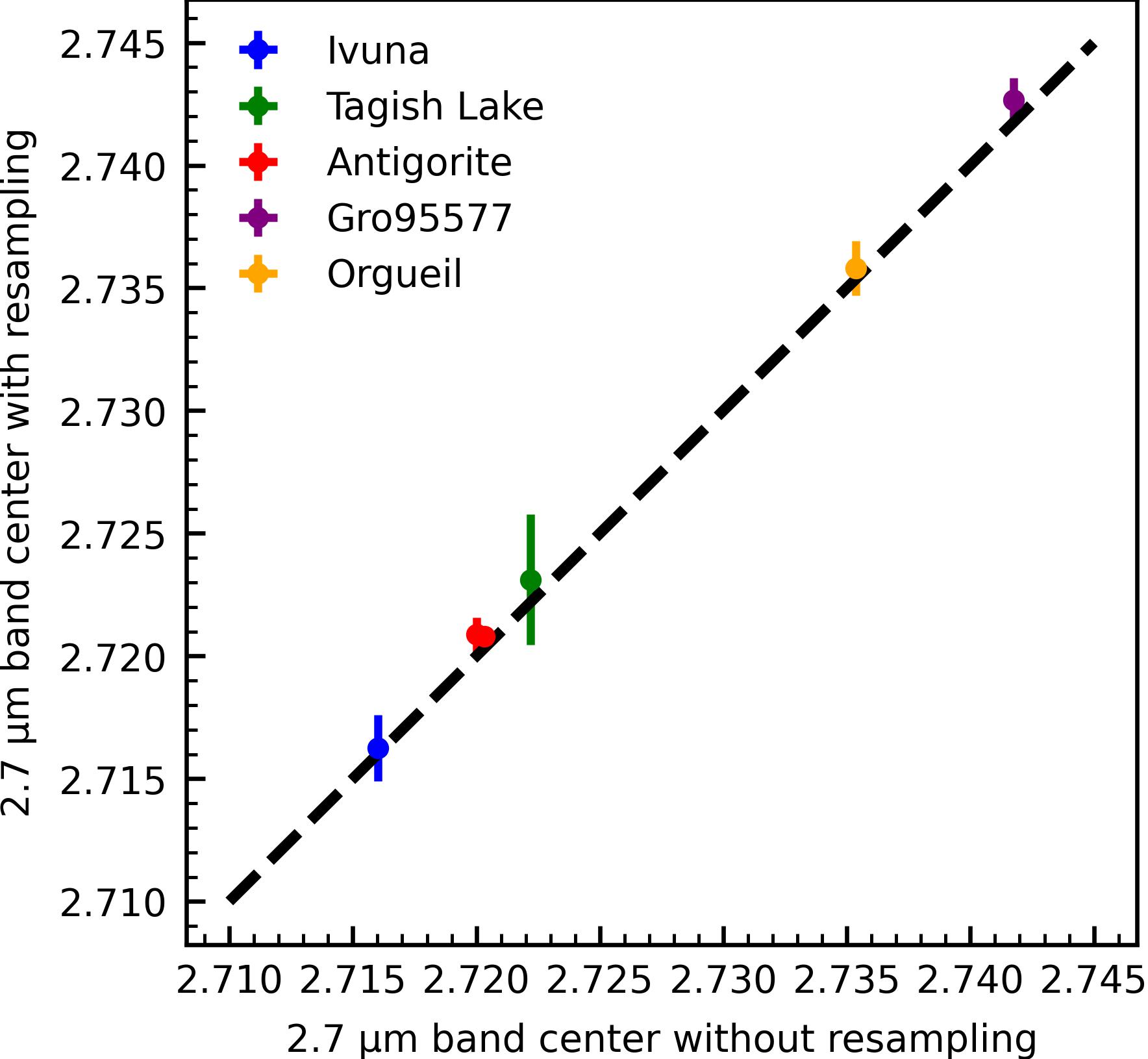}
\caption{The influence of VIR spectral sampling on the \SI{2.7}{\um} band center. The x-axis is the fitted \SI{2.7}{\um} band center with the original laboratory measured spectra and the y-axis represents the band center fitted from the spectra with the VIR spectral sampling. The dashed line is the 1:1 line. The errorbars are the uncertainty determined with the different SNR values ranging from \num{100} to \num{130}.
\label{fig:spectral sampling}}
\end{figure}

We compared the \SI{2.7}{\um} band center of different carbonaceous chondrites and antigorite samples measured in the laboratory with different spectral sampling. The Ivuna and Tagish Lake were from \cite{2020JGRE..12506606K}. Antigorite was from RELAB (sample ID: AT-TXH-006 and AT-TXH-007). GRO9557 and Orgueil were from \cite{2020Icar..34813826P}. All samples were measured in a dry ambient atmosphere or heated before measurement to minimize adsorbed water. We resampled these laboratory spectra with VIR wavelength sampling. To account for the higher noise in VIR data compared to laboratory spectra, we added Gaussian noise to the resampled laboratory spectra at a level of signal-to-noise ratio (SNR) of \num{100}\text{--}\num{130}, comparable to a primary estimation from VIR of $\sim$\num{120}. We then used the same Gaussian fit procedure to determine the \SI{2.7}{\um} band center from both the original laboratory spectra and after spectral resampling. The result showed that the band centers retrieved from the resampled and noise-added spectra are consistent with those retrieved from the original spectra within an uncertainty of $<$\SI{2}{nm} (Fig. \ref{fig:spectral sampling}). This test demonstrates that the VIR spectral sampling of \SI{9.8}{\nm} is not a limiting factor to estimate the \SI{2.7}{\um} band center to a precision of a few \si{nm}.

\section{Sensitivity analysis of the \SI{2.7}{\um} band center to other possible minor phases} \label{sec:sensitivity analysis}

To check whether minor amounts of \ce{H2O} ice and salts that are undetectable in VIR data affect the \SI{2.7}{\um} band center to a level of a few nm, we conducted a spectral mixing test. Here, we adopted a linear spectral mixing model, as used in \cite{2016Sci...353.3010C}, which is adequate for our purpose of assessing how mixtures affect the \SI{2.7}{\um} band center. The Ceres average and \ce{H2O} ice spectra were from \cite{2016Sci...353.3010C}. For salts, we used the spectrum of a brine mixture of ammonium chloride (\ce{NH4Cl}), natrite (\ce{Na2CO3}), and hydrohalite (\ce{NaCl*2H2O}), called Solution 3 (Sol. 3) from \cite{2019Icar..320..150T}. This is the same spectrum used by \cite{2020NatAs...4..786D} to identify salts from the Cerealia tholus on Ceres. We linearly mixed the \ce{H2O} ice and salts spectra with the Ceres average spectrum, respectively, then used the same Gaussian fit as in our analysis to determine the \SI{2.7}{\um} band center for the Ceres average and mixture spectra.

\begin{figure}[ht!]
\plotone{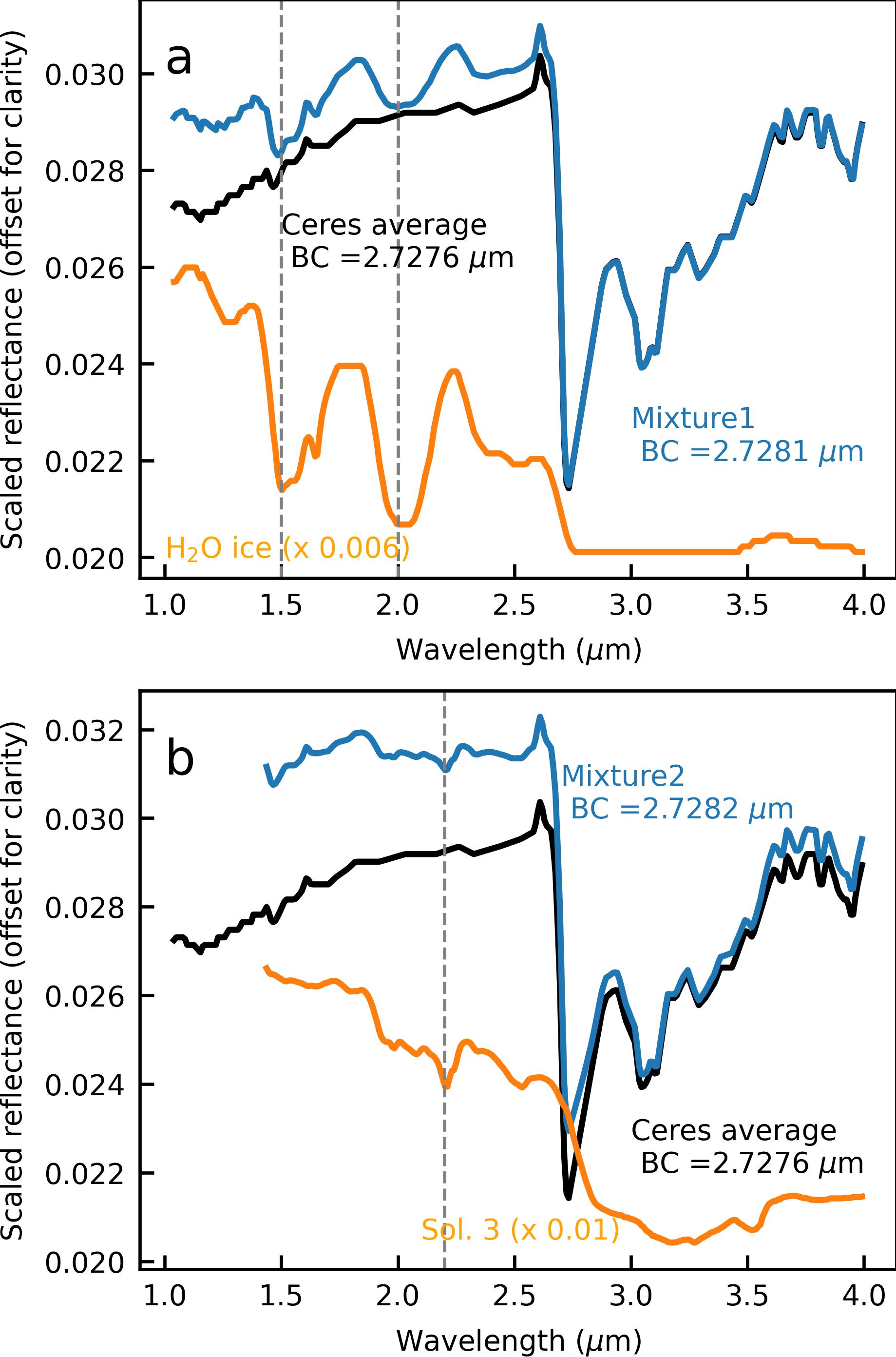}
\caption{Sensitivity analysis of the \SI{2.7}{\um} band center to minor amounts of \ce{H2O} ice and salts. In each panel, the black spectrum is the Ceres average spectrum from \cite{2016Sci...353.3010C}; the orange spectrum corresponds to (a) the \ce{H2O} ice spectrum from \cite{2016Sci...353.3010C} and (b) the brine spectrum (Sol. 3) from \cite{2019Icar..320..150T}; the blue spectrum represents a mixture of the Ceres average with (a) \ce{H2O} ice or (b) salts. The band centers of the Ceres average and the mixtures are noted. The vertical dashed lines illustrate the diagnostic features of \ce{H2O} ice and salts.
\label{fig:sensitivity analysis}}
\end{figure}

Fig. \ref{fig:sensitivity analysis}a shows an example of a mixture composed of \SI{99.8}{\%} Ceres average and \SI{0.2}{\%} \ce{H2O} ice. The \num{1.5} and \SI{2.0}{\um} absorptions of \ce{H2O} ice are still detectable, whereas the \SI{2.7}{\um} band center of the mixture shifts by less than \SI{1}{\nm} compared to that of the Ceres average. Fig. \ref{fig:sensitivity analysis}b shows an example of a mixture of \SI{99.5}{\%} Ceres average and \SI{0.5}{\%} Sol. 3. The \SI{2.2}{\um} absorption feature is slightly visible, while the resulting shift of the \SI{2.7}{\um} band center also remains within \SI{1}{\nm}. Therefore, minor amounts of \ce{H2O} ice or salts at undetectable levels of abundance should cause the \SI{2.7}{\um} band shift of \textless\SI{1}{nm} and have a negligible effect on the position of the \SI{2.7}{\um} band center for our purpose.

\section{Spectra of blue-shifted region} \label{sec:blue_shifted_spectra}

\begin{figure*}[p]
\plotone{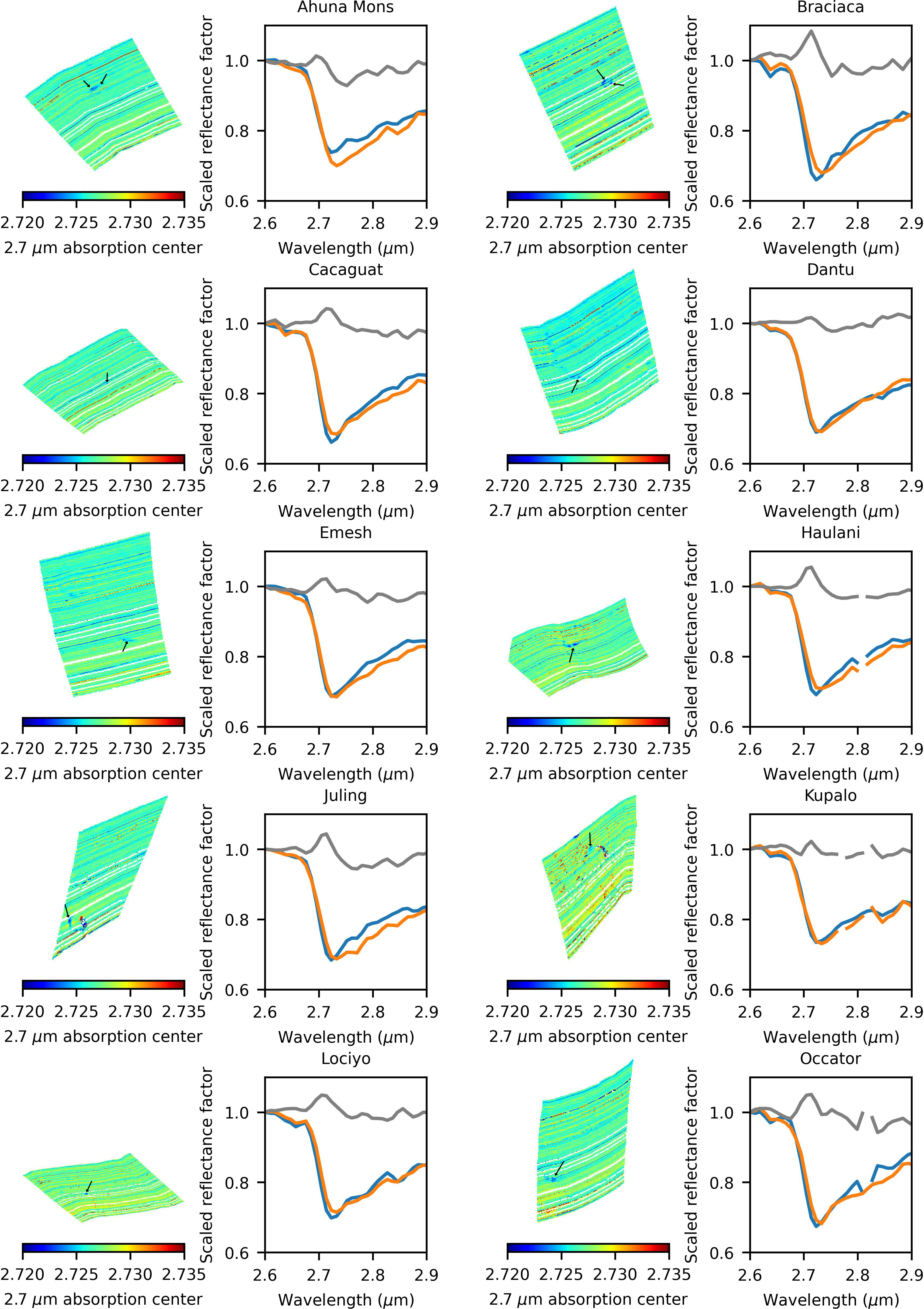}
\caption{\SI{2.7}{\um} band center mapping of different blue-shifted regions and the corresponding spectra. The black arrows illustrate the location of blue-shifted regions. The blue and orange spectra represent the blue-shifted regions and adjacent background, respectively. The gray spectra are ratioed spectra of blue-shifted and adjacent regions.
\label{fig:all spectra}}
\end{figure*}

\begin{figure*}[p]
\plotone{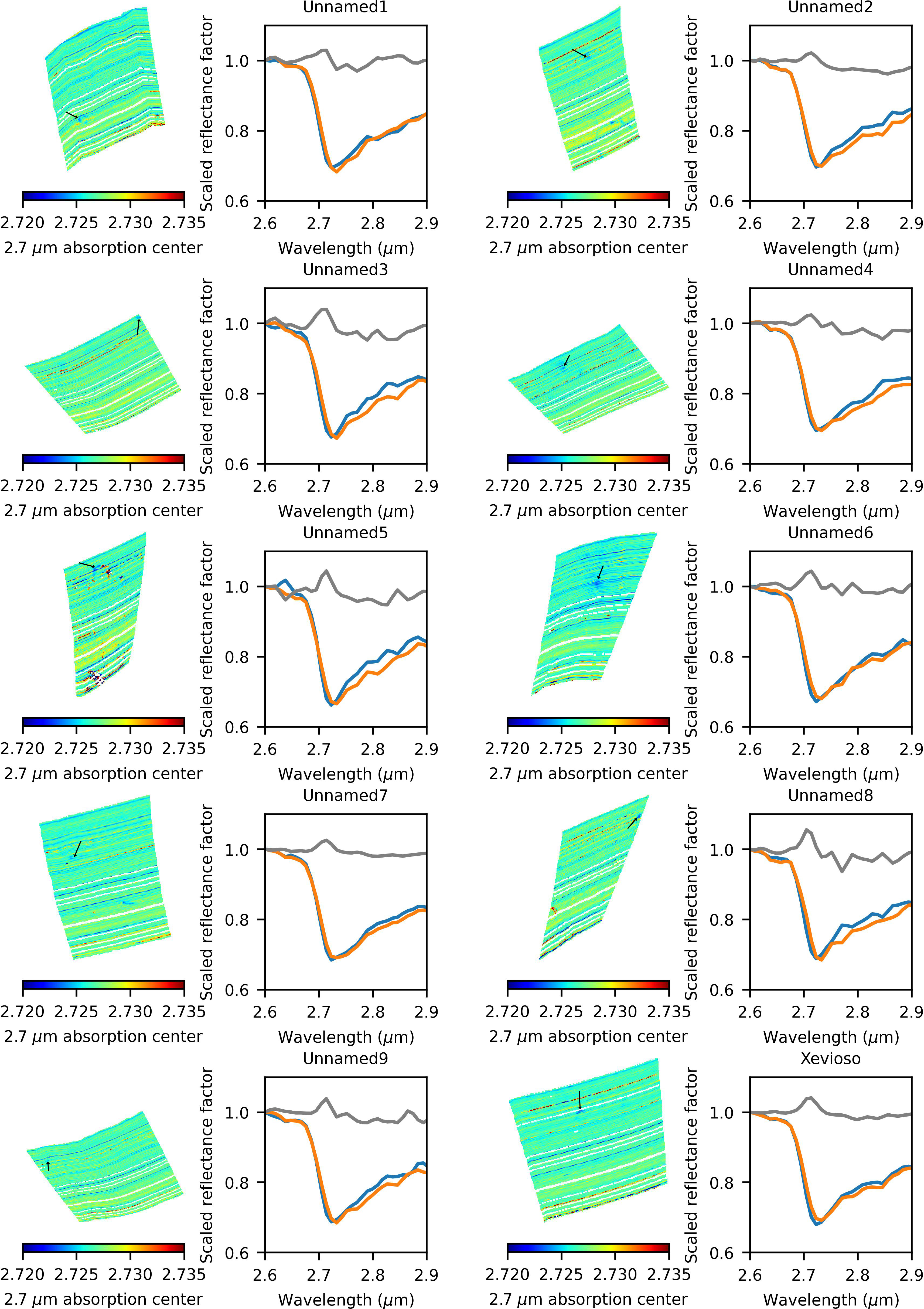}\setcounter{figure}{\value{figure}-1}
\caption{\SI{2.7}{\um} band center mapping of different blue-shifted regions and the corresponding spectra. The black arrows illustrate the location of blue-shifted regions. The blue and orange spectra represent the blue-shifted regions and adjacent background, respectively. The gray spectra are ratioed spectra of blue-shifted and adjacent regions. (Continued).}
\end{figure*}

\section{Spatial relationship between blue-shifted regions and other composition} \label{sec:distribution_with_others}

\begin{figure*}[ht!]
\plotone{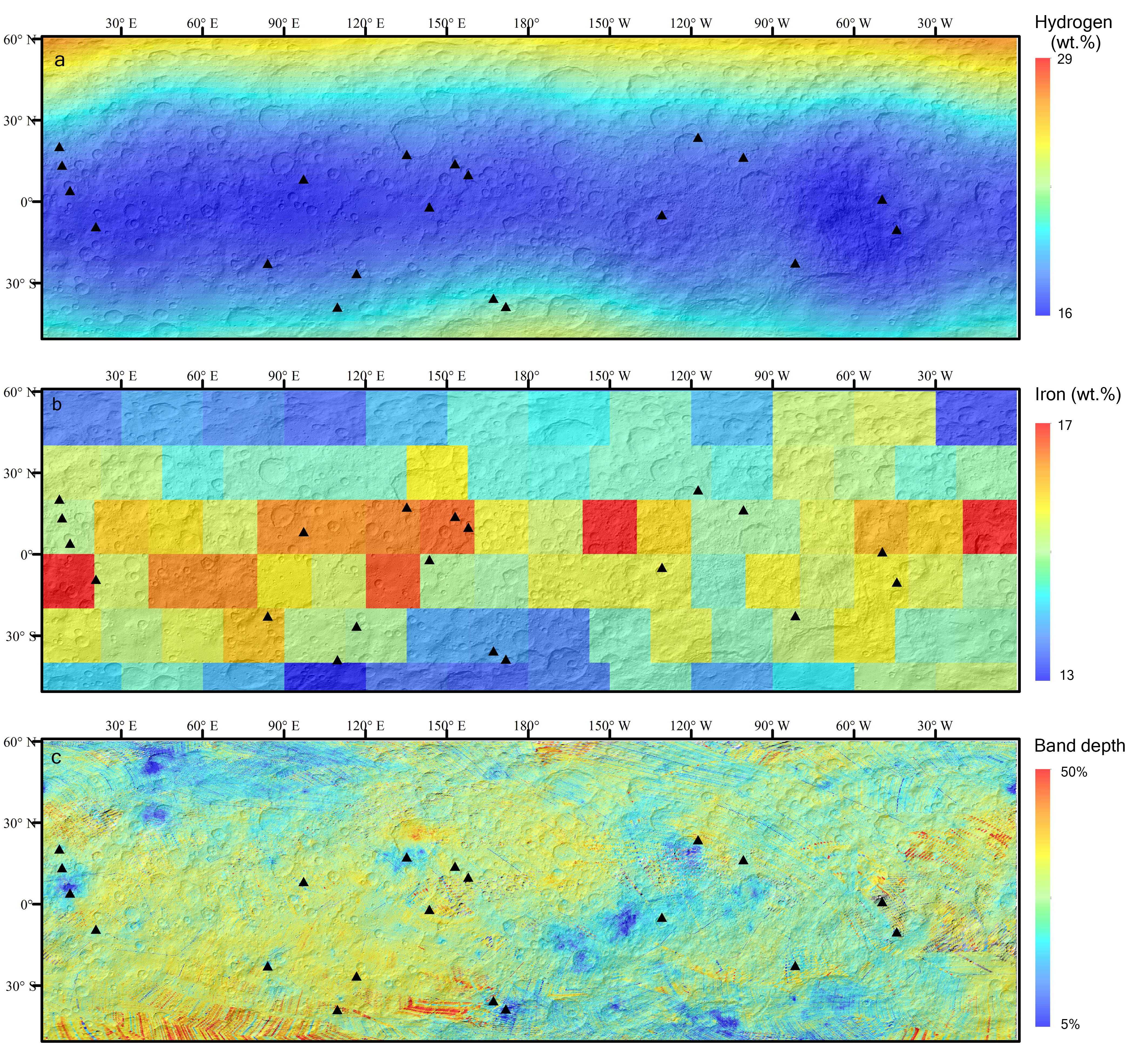}
\caption{\added{Spatial distribution of the blue-shifted regions overlaid on maps of (a) hydrogen abundance, (b) iron abundance, and (c) the \SI{2.7}{\um} band depth of phyllosilicates on Ceres. The blue-shifted regions are marked with the black triangles. These compositional maps are publicly available from the PDS archive at \url{https://arcnav.psi.edu/urn:nasa:pds:context:investigation:mission.dawn_mission_to_vesta_and_ceres/more}}
\label{fig:distribution_others}}
\end{figure*}

\section{blue-shifted regions}

\tabletypesize{\footnotesize}
\begin{deluxetable*}{ccccccccc}
\tablewidth{0pt}
\tablecaption{Location of blue-shifted regions and the corresponding spectral parameters \label{table:blue_shifted}}
\tablehead{
\colhead{Unit} & \colhead{Location} & \colhead{Distribution} & \colhead{BC2.7} & \colhead{VNIR slope} & \colhead{BD2.7} & \colhead{\shortstack{Reflectance \\(0.44 $\mu$m)}} & \colhead{BD3.1} & \colhead{BD3.9} 
}
\startdata
Ahuna & 315.64$^{\circ}$E, 10.20$^{\circ}$S & Mons flank 
& \shortstack{2.7255 \\ ($\pm$0.0020)}
& \shortstack{0.919 \\ ($\pm$0.020)}
& \shortstack{0.206 \\ ($\pm$0.018)}
& \shortstack{0.0418 \\ ($\pm$0.0032)} 
& \shortstack{0.0535 \\ ($\pm$0.0086)} 
& \shortstack{0.1128 \\ ($\pm$0.0072)} \\
Braciaca1 & 84.54$^{\circ}$E, 22.99$^{\circ}$S & Crater wall 
& \shortstack{2.7231 \\ ($\pm$0.0012)}
& \shortstack{1.093 \\ ($\pm$0.058)}
& \shortstack{0.305 \\ ($\pm$0.009)}
& \shortstack{0.0291 \\ ($\pm$0.0042)} 
& \shortstack{0.0813 \\ ($\pm$0.0063)} 
& \shortstack{0.0555 \\ ($\pm$0.0130)} \\
Braciaca2 & 84.05$^{\circ}$E, 22.71$^{\circ}$S & Crater wall 
& \shortstack{2.7237 \\ ($\pm$0.0013)}
& \shortstack{1.018 \\ ($\pm$0.015)}
& \shortstack{0.277 \\ ($\pm$0.006)}
& \shortstack{0.0309 \\ ($\pm$0.0019)} 
& \shortstack{0.0882 \\ ($\pm$0.0122)} 
& \shortstack{0.0539 \\ ($\pm$0.0144)} \\
Cacaguat & 143.71$^{\circ}$E, 1.78$^{\circ}$S & Crater wall 
& \shortstack{2.7245 \\ ($\pm$0.0005)}
& \shortstack{1.060 \\ ($\pm$0.027)}
& \shortstack{0.292 \\ ($\pm$0.009)}
& \shortstack{0.0299 \\ ($\pm$0.0011)} 
& \shortstack{0.0948 \\ ($\pm$0.0067)} 
& \shortstack{0.0845 \\ ($\pm$0.0110)} \\
Dantu & 135.24$^{\circ}$E, 17.54$^{\circ}$N & Crater wall 
& \shortstack{2.7248 \\ ($\pm$0.0012)}
& \shortstack{1.025 \\ ($\pm$0.050)}
& \shortstack{0.280 \\ ($\pm$0.008)}
& \shortstack{0.0331 \\ ($\pm$0.0144)} 
& \shortstack{0.0826 \\ ($\pm$0.0106)} 
& \shortstack{0.1572 \\ ($\pm$0.0252)} \\
Emesh & 157.71$^{\circ}$E, 10.14$^{\circ}$N & Crater wall 
& \shortstack{2.7246 \\ ($\pm$0.0012)}
& \shortstack{1.049 \\ ($\pm$0.026)}
& \shortstack{0.276 \\ ($\pm$0.007)}
& \shortstack{0.0324 \\ ($\pm$0.0015)} 
& \shortstack{0.0937 \\ ($\pm$0.0063)} 
& \shortstack{0.0797 \\ ($\pm$0.0135)} \\
Haulani1 & 11.48$^{\circ}$E, 4.23$^{\circ}$N & Crater wall 
& \shortstack{2.7233 \\ ($\pm$0.0016)}
& \shortstack{1.027 \\ ($\pm$0.047)}
& \shortstack{0.267 \\ ($\pm$0.017)}
& \shortstack{0.0316 \\ ($\pm$0.0047)} 
& \shortstack{0.0756 \\ ($\pm$0.0069)} 
& \shortstack{0.0839 \\ ($\pm$0.0171)} \\
Haulani2 & 10.40$^{\circ}$E, 3.97$^{\circ}$N & Crater wall 
& \shortstack{2.7235 \\ ($\pm$0.0011)}
& \shortstack{1.016 \\ ($\pm$0.025)}
& \shortstack{0.269 \\ ($\pm$0.007)}
& \shortstack{0.0317 \\ ($\pm$0.0025)} 
& \shortstack{0.0734 \\ ($\pm$0.0078)} 
& \shortstack{0.0786 \\ ($\pm$0.0056)} \\
Haulani3 & 9.47$^{\circ}$E, 4.69$^{\circ}$N & Crater wall 
& \shortstack{2.7244 \\ ($\pm$0.0009)}
& \shortstack{1.007 \\ ($\pm$0.019)}
& \shortstack{0.263 \\ ($\pm$0.011)}
& \shortstack{0.0312 \\ ($\pm$0.0021)} 
& \shortstack{0.0759 \\ ($\pm$0.0038)} 
& \shortstack{0.0381 \\ ($\pm$0.0049)} \\
Haulani4 & 10.41$^{\circ}$E, 7.48$^{\circ}$N & Crater wall 
& \shortstack{2.7238 \\ ($\pm$0.0011)}
& \shortstack{0.991 \\ ($\pm$0.024)}
& \shortstack{0.271 \\ ($\pm$0.006)}
& \shortstack{0.0345 \\ ($\pm$0.0015)} 
& \shortstack{0.0659 \\ ($\pm$0.0075)} 
& \shortstack{0.0835 \\ ($\pm$0.0154)} \\
Haulani5 & 11.39$^{\circ}$E, 7.54$^{\circ}$N & Crater wall 
& \shortstack{2.7245 \\ ($\pm$0.0013)}
& \shortstack{0.972 \\ ($\pm$0.046)}
& \shortstack{0.267 \\ ($\pm$0.010)}
& \shortstack{0.0346 \\ ($\pm$0.0029)} 
& \shortstack{0.0735 \\ ($\pm$0.0078)} 
& \shortstack{0.0827 \\ ($\pm$0.0075)} \\
Haulani6 & 12.07$^{\circ}$E, 7.15$^{\circ}$N & Crater wall 
& \shortstack{2.7251 \\ ($\pm$0.0014)}
& \shortstack{0.999 \\ ($\pm$0.053)}
& \shortstack{0.259 \\ ($\pm$0.007)}
& \shortstack{0.0338 \\ ($\pm$0.0035)} 
& \shortstack{0.0666 \\ ($\pm$0.0090)} 
& \shortstack{0.1177 \\ ($\pm$0.0265)} \\
Juling & 167.24$^{\circ}$E, 35.46$^{\circ}$S & Crater wall 
& \shortstack{2.7243 \\ ($\pm$0.0016)}
& \shortstack{0.972 \\ ($\pm$0.049)}
& \shortstack{0.278 \\ ($\pm$0.017)}
& \shortstack{0.0384 \\ ($\pm$0.0062)} 
& \shortstack{0.0879 \\ ($\pm$0.0082)} 
& \shortstack{0.1464 \\ ($\pm$0.0321)} \\
Kupalo & 171.81$^{\circ}$E, 38.59$^{\circ}$S & Crater wall 
& \shortstack{2.7253 \\ ($\pm$0.0012)}
& \shortstack{0.877 \\ ($\pm$0.058)}
& \shortstack{0.219 \\ ($\pm$0.012)}
& \shortstack{0.0547 \\ ($\pm$0.0105)} 
& \shortstack{0.0633 \\ ($\pm$0.0058)} 
& \shortstack{0.1478 \\ ($\pm$0.0146)} \\
Lociyo & 229.24$^{\circ}$E, 4.68$^{\circ}$S & Crater wall 
& \shortstack{2.7246 \\ ($\pm$0.0009)}
& \shortstack{1.036 \\ ($\pm$0.019)}
& \shortstack{0.278 \\ ($\pm$0.009)}
& \shortstack{0.0283 \\ ($\pm$0.0012)} 
& \shortstack{0.0586 \\ ($\pm$0.0035)} 
& \shortstack{0.0428 \\ ($\pm$0.0116)} \\
Occator1 & 237.41$^{\circ}$E, 15.37$^{\circ}$N & Crater wall 
& \shortstack{2.7249 \\ ($\pm$0.0007)}
& \shortstack{1.021 \\ ($\pm$0.047)}
& \shortstack{0.270 \\ ($\pm$0.006)}
& \shortstack{0.0292 \\ ($\pm$0.0023)} 
& \shortstack{0.0732 \\ ($\pm$0.0040)} 
& \shortstack{0.1210 \\ ($\pm$0.0110)} \\
Occator2 & 235.48$^{\circ}$E, 16.11$^{\circ}$N & Crater wall 
& \shortstack{2.7244 \\ ($\pm$0.0006)}
& \shortstack{0.980 \\ ($\pm$0.040)}
& \shortstack{0.258 \\ ($\pm$0.009)}
& \shortstack{0.0307 \\ ($\pm$0.0018)} 
& \shortstack{0.0712 \\ ($\pm$0.0045)} 
& \shortstack{0.1224 \\ ($\pm$0.0137)} \\
Occator3 & 234.02$^{\circ}$E, 19.40$^{\circ}$N & Crater wall 
& \shortstack{2.7242 \\ ($\pm$0.0007)}
& \shortstack{1.025 \\ ($\pm$0.037)}
& \shortstack{0.277 \\ ($\pm$0.007)}
& \shortstack{0.0301 \\ ($\pm$0.0029)} 
& \shortstack{0.0774 \\ ($\pm$0.0037)} 
& \shortstack{0.1134 \\ ($\pm$0.0086)} \\
Occator4 & 241.81$^{\circ}$E, 23.95$^{\circ}$N & Crater wall 
& \shortstack{2.7239 \\ ($\pm$0.0005)}
& \shortstack{1.069 \\ ($\pm$0.016)}
& \shortstack{0.297 \\ ($\pm$0.005)}
& \shortstack{0.0260 \\ ($\pm$0.0006)} 
& \shortstack{0.0925 \\ ($\pm$0.0149)} 
& \shortstack{0.1003 \\ ($\pm$0.0072)} \\
Occator5 & 242.49$^{\circ}$E, 23.94$^{\circ}$N & Crater wall 
& \shortstack{2.7245 \\ ($\pm$0.0007)}
& \shortstack{1.040 \\ ($\pm$0.031)}
& \shortstack{0.297 \\ ($\pm$0.007)}
& \shortstack{0.0276 \\ ($\pm$0.0011)} 
& \shortstack{0.0819 \\ ($\pm$0.0135)} 
& \shortstack{0.1021 \\ ($\pm$0.0078)} \\
Unnamed1 & 7.31$^{\circ}$E, 20.46$^{\circ}$N & Crater wall 
& \shortstack{2.7245 \\ ($\pm$0.0006)}
& \shortstack{1.044 \\ ($\pm$0.024)}
& \shortstack{0.277 \\ ($\pm$0.006)}
& \shortstack{0.0306 \\ ($\pm$0.0014)} 
& \shortstack{0.0744 \\ ($\pm$0.0048)} 
& \shortstack{0.1155 \\ ($\pm$0.0213)} \\
Unnamed2 & 8.28$^{\circ}$E, 13.51$^{\circ}$N & Small crater 
& \shortstack{2.7245 \\ ($\pm$0.0004)}
& \shortstack{0.996 \\ ($\pm$0.017)}
& \shortstack{0.265 \\ ($\pm$0.005)}
& \shortstack{0.0337 \\ ($\pm$0.0036)} 
& \shortstack{0.0833 \\ ($\pm$0.0071)} 
& \shortstack{0.0461 \\ ($\pm$0.0146)} \\
Unnamed3 & 20.67$^{\circ}$E, 9.17$^{\circ}$S & Crater wall 
& \shortstack{2.7247 \\ ($\pm$0.0011)}
& \shortstack{1.019 \\ ($\pm$0.025)}
& \shortstack{0.274 \\ ($\pm$0.005)}
& \shortstack{0.0295 \\ ($\pm$0.0020)} 
& \shortstack{0.0666 \\ ($\pm$0.0049)} 
& \shortstack{0.0502 \\ ($\pm$0.0099)} \\
Unnamed4 & 97.30$^{\circ}$E, 8.45$^{\circ}$N & Crater wall 
& \shortstack{2.7244 \\ ($\pm$0.0011)}
& \shortstack{1.063 \\ ($\pm$0.015)}
& \shortstack{0.274 \\ ($\pm$0.005)}
& \shortstack{0.0356 \\ ($\pm$0.0023)} 
& \shortstack{0.0739 \\ ($\pm$0.0032)} 
& \shortstack{0.0793 \\ ($\pm$0.0098)} \\
Unnamed5 & 109.70$^{\circ}$E, 38.80$^{\circ}$S & Crater wall 
& \shortstack{2.7235 \\ ($\pm$0.0011)}
& \shortstack{1.053 \\ ($\pm$0.014)}
& \shortstack{0.306 \\ ($\pm$0.006)}
& \shortstack{0.0316 \\ ($\pm$0.0027)} 
& \shortstack{0.0959 \\ ($\pm$0.0063)} 
& \shortstack{0.0967 \\ ($\pm$0.0193)} \\
Unnamed6 & 116.80$^{\circ}$E, 26.36$^{\circ}$S & Small crater 
& \shortstack{2.7231 \\ ($\pm$0.0011)}
& \shortstack{1.097 \\ ($\pm$0.016)}
& \shortstack{0.295 \\ ($\pm$0.008)}
& \shortstack{0.0313 \\ ($\pm$0.0020)} 
& \shortstack{0.0997 \\ ($\pm$0.0066)} 
& \shortstack{0.1148 \\ ($\pm$0.0118)} \\
Unnamed7 & 153.10$^{\circ}$E, 14.01$^{\circ}$N & Crater wall 
& \shortstack{2.7251 \\ ($\pm$0.0010)}
& \shortstack{1.012 \\ ($\pm$0.025)}
& \shortstack{0.275 \\ ($\pm$0.004)}
& \shortstack{0.0307 \\ ($\pm$0.0018)} 
& \shortstack{0.0918 \\ ($\pm$0.0049)} 
& \shortstack{0.1349 \\ ($\pm$0.0162)} \\
Unnamed8 & 278.44$^{\circ}$E, 22.38$^{\circ}$S & Crater wall 
& \shortstack{2.7241 \\ ($\pm$0.0014)}
& \shortstack{1.003 \\ ($\pm$0.057)}
& \shortstack{0.271 \\ ($\pm$0.012)}
& \shortstack{0.0319 \\ ($\pm$0.0020)} 
& \shortstack{0.0838 \\ ($\pm$0.0066)} 
& \shortstack{0.0728 \\ ($\pm$0.0089)} \\
Unnamed9 & 259.33$^{\circ}$E, 16.44$^{\circ}$N & Small crater 
& \shortstack{2.7250 \\ ($\pm$0.0003)}
& \shortstack{1.065 \\ ($\pm$0.011)}
& \shortstack{0.275 \\ ($\pm$0.003)}
& \shortstack{0.0288 \\ ($\pm$0.0012)} 
& \shortstack{0.0746 \\ ($\pm$0.0088)} 
& \shortstack{0.1360 \\ ($\pm$0.0187)} \\
Xevioso & 310.49$^{\circ}$E, 1.07$^{\circ}$N & Crater wall 
& \shortstack{2.7239 \\ ($\pm$0.0009)}
& \shortstack{1.018 \\ ($\pm$0.040)}
& \shortstack{0.279 \\ ($\pm$0.005)}
& \shortstack{0.0342 \\ ($\pm$0.0018)} 
& \shortstack{0.0823 \\ ($\pm$0.0055)} 
& \shortstack{0.0994 \\ ($\pm$0.0117)} \\
\enddata
\tablecomments{Some crater walls contain more than one sporadically distributed blue-shifted region. We count these regions separately.}
\end{deluxetable*}

\section{Absolute age of some typical regions} \label{sec:age}

\tabletypesize{\footnotesize}
\begin{deluxetable*}{cccccccc}
\tablewidth{0pt}
\tablecaption{LDM age and spectral parameters for some typical units \label{table:age}}
\tablehead{
\colhead{Unit} & \colhead{Location} & \colhead{LDM (Ma)}\tablenotemark{a} & \colhead{VNIR slope} & \colhead{BD2.7} & \colhead{\shortstack{Reflectance \\(0.44 $\mu$m)}} & \colhead{BD3.1} & \colhead{BD3.9} 
}
\startdata
Achita & 65.9$^{\circ}$E, 25.8$^{\circ}$N & 570 ($\pm$60) & 1.022 ($\pm$0.009) & 0.269 ($\pm$0.007) & 0.0322 ($\pm$0.0022) & 0.072 ($\pm$0.006) & 0.110 ($\pm$0.011) \\
Ahuna & 316.2$^{\circ}$E, 10.5$^{\circ}$S & 70 ($\pm$20) & 0.984 ($\pm$0.014) & 0.254 ($\pm$0.009) & 0.0325 ($\pm$0.0015) & 0.060	($\pm$0.008) & 0.057 ($\pm$0.017) \\
Azacca & 218.4$^{\circ}$E, 6.7$^{\circ}$S & 75.9 ($\pm$10) & 0.983 ($\pm$0.008) & 0.259 ($\pm$0.014) & 0.0324 ($\pm$0.0012) & 0.055 ($\pm$0.008) & 0.029 ($\pm$0.011) \\
Cacaguat & 143.6$^{\circ}$E, 1.2$^{\circ}$S & 1.3 ($\pm$0.79) & 0.960 ($\pm$0.012) & 0.262 ($\pm$0.009) & 0.0363 ($\pm$0.0014) & 0.084 ($\pm$0.005) & 0.044 ($\pm$0.013) \\
Centeotl & 141.2$^{\circ}$E, 18.9$^{\circ}$N & 4.2 ($\pm$3) & 0.927 ($\pm$0.009) & 0.237 ($\pm$0.010) & 0.0364 ($\pm$0.0010) & 0.060 ($\pm$0.006) & 0.114 ($\pm$0.010) \\
Coniraya & 65.7$^{\circ}$E, 39.9$^{\circ}$N & 1300 ($\pm$300) & 1.027 ($\pm$0.012) & 0.263 ($\pm$0.012) & 0.0306 ($\pm$0.0026) & 0.066 ($\pm$0.013) & 0.064 ($\pm$0.010) \\
Dantu & 138.2$^{\circ}$E, 24.3$^{\circ}$N & 111 ($\pm$39) & 1.031 ($\pm$0.010) & 0.284 ($\pm$0.004) & 0.0348 ($\pm$0.0013) & 0.095 ($\pm$0.006) & 0.128 ($\pm$0.009) \\
Ernutet & 45.5$^{\circ}$E, 52.9$^{\circ}$N & 1600 ($\pm$200) & 1.044 ($\pm$0.012) & 0.252 ($\pm$0.006) & 0.0317 ($\pm$0.0016) & 0.067 ($\pm$0.005) & 0.079 ($\pm$0.008) \\
Gaue & 86.2$^{\circ}$E, 30.8$^{\circ}$N & 260 ($\pm$30) & 1.024 ($\pm$0.008) & 0.266 ($\pm$0.004) & 0.0322 ($\pm$0.0012) & 0.076 ($\pm$0.006) & 0.102 ($\pm$0.011) \\
Haulani & 10.8$^{\circ}$E, 5.8$^{\circ}$N & 2.7 ($\pm$0.7) & 0.874 ($\pm$0.006) & 0.229 ($\pm$0.029) & 0.0423 ($\pm$0.0015) & 0.032 ($\pm$0.008) & 0.082 ($\pm$0.010) \\
Ikapati1\tablenotemark{b} & 44.4$^{\circ}$E, 33.8$^{\circ}$N & 19.2 ($\pm$2.2) & 0.966 ($\pm$0.005) & 0.243 ($\pm$0.007) & 0.0348 ($\pm$0.0010) & 0.055 ($\pm$0.006) & 0.047 ($\pm$0.003) \\
Ikapati3 & 40.9$^{\circ}$E, 35.0$^{\circ}$N & 42 ($\pm$8.3) & 0.937 ($\pm$0.012) & 0.255 ($\pm$0.007) & 0.0361 ($\pm$0.0024) & 0.050	($\pm$0.011) & 0.116 ($\pm$0.007) \\
Juling & 168.5$^{\circ}$E, 35.9$^{\circ}$S & $<$ 2.5 & 0.924 ($\pm$0.025) & 0.232 ($\pm$0.007) & 0.0382 ($\pm$0.0034) & 0.060 ($\pm$0.006) & 0.139 ($\pm$0.010) \\
Kerwan & 124.0$^{\circ}$E, 10.8$^{\circ}$S & 1300 ($\pm$160) & 1.065 ($\pm$0.014) & 0.281 ($\pm$0.007) & 0.0320 ($\pm$0.0015) & 0.084 ($\pm$0.006) & 0.051 ($\pm$0.019) \\
Kupalo & 173.2$^{\circ}$E, 39.4$^{\circ}$S & $<$ 4.5 & 0.932 ($\pm$0.020) & 0.245 ($\pm$0.014) & 0.0399 ($\pm$0.0036) & 0.064 ($\pm$0.010) & 0.091 ($\pm$0.010) \\
Liber & 37.8$^{\circ}$E, 42.6$^{\circ}$N & 440 ($\pm$60) & 1.031 ($\pm$0.011) & 0.263 ($\pm$0.006) & 0.0303 ($\pm$0.0016) & 0.069 ($\pm$0.005) & 0.072 ($\pm$0.012) \\
Messor & 233.7$^{\circ}$E, 49.9$^{\circ}$N & 64.5 ($\pm$2.6) & 1.011 ($\pm$0.012) & 0.268 ($\pm$0.005) & 0.0317 ($\pm$0.0018) & 0.051 ($\pm$0.017) & 0.077 ($\pm$0.009) \\
Omonga & 71.7$^{\circ}$E, 58.0$^{\circ}$N & 970 ($\pm$70) & 1.024 ($\pm$0.013) & 0.264 ($\pm$0.010) & 0.0308 ($\pm$0.0019) & 0.057 ($\pm$0.008) & 0.056 ($\pm$0.010) \\
Rao & 119.0$^{\circ}$E, 8.1$^{\circ}$N & 33.1 ($\pm$2.5) & 1.004 ($\pm$0.008) & 0.278 ($\pm$0.005) & 0.0387 ($\pm$0.0023) & 0.098 ($\pm$0.005) & 0.112 ($\pm$0.029) \\
Sintana & 46.2$^{\circ}$E, 48.1$^{\circ}$S & 310 ($\pm$40) & 1.060 ($\pm$0.017) & 0.280 ($\pm$0.011) & 0.0311 ($\pm$0.0023) & 0.075 ($\pm$0.007) & 0.102 ($\pm$0.023) \\
Tupo & 88.4$^{\circ}$E, 32.3$^{\circ}$S & 49 ($\pm$8) & 0.996 ($\pm$0.008) & 0.262 ($\pm$0.005) & 0.0329 ($\pm$0.0015) & 0.073 ($\pm$0.008) & 0.117 ($\pm$0.010) \\
Unnamed & 247$^{\circ}$E, 39$^{\circ}$N & 906 ($\pm$130) & 1.038 ($\pm$0.018) & 0.279 ($\pm$0.005) & 0.0294 ($\pm$0.0018) & 0.075 ($\pm$0.004) & 0.149 ($\pm$0.009) \\
Urvara & 249.2$^{\circ}$E, 46.6$^{\circ}$S & 134 ($\pm$8) & 1.039 ($\pm$0.010) & 0.261 ($\pm$0.009) & 0.0308 ($\pm$0.0012) & 0.066 ($\pm$0.011) & 0.118 ($\pm$0.007) \\
Yalode & 292.5$^{\circ}$E, 42.6$^{\circ}$S & 1100 ($\pm$450) & 1.045 ($\pm$0.009) & 0.261 ($\pm$0.009) & 0.0318 ($\pm$0.0012) & 0.073 ($\pm$0.006) & 0.102 ($\pm$0.010) \\
\enddata
\tablenotetext{a}{LDM data used in this study are from \cite{2016GeoRL..4311987S, 2017GeoRL..44.6570S, 2019Icar..318...56S}}
\tablenotetext{b}{The numbering is consistent with that used in \cite{2016GeoRL..4311987S}}
\end{deluxetable*}


\bibliography{sample701}{}
\bibliographystyle{aasjournalv7}


\end{CJK*}

\end{document}